\documentclass[12pt]{article}
\usepackage{amsmath}
\usepackage{graphicx}
\usepackage{enumerate}
\usepackage{natbib}
\usepackage{url} 
\usepackage{threeparttable, multirow, array}
\usepackage{bm}
\newcommand{\blind}{1}

\begin{document}

\def\spacingset#1{\renewcommand{\baselinestretch}%
{#1}\small\normalsize} \spacingset{1}


\if1\blind
{
  \title{\bf Addressing Detection Limits with Semiparametric Cumulative Probability Models}
  \author{Yuqi Tian\thanks{
    The authors gratefully acknowledge CCASAnet investigators for providing data for the HIV study. This study was supported by funding from the U.S. National Institutes of Health, grants \textit{R01 AI093234} and \textit{U01 AI069923}. }\hspace{.2cm}\\
    Department of Biostatistics, Vanderbilt University,\\
    Chun Li \\
    Department of Population and Public Health Sciences, \\University of Southern California,\\
    Shengxin Tu \\
    Department of Biostatistics, Vanderbilt University,\\
    Nathan T. James \\
    Department of Biostatistics, Vanderbilt University,\\
    Frank E. Harrell \\
    Department of Biostatistics, Vanderbilt University,\\
    and \\
    Bryan E. Shepherd\\
    Department of Biostatistics, Vanderbilt University}
  \maketitle
} \fi

\if0\blind
{
  \bigskip
  \bigskip
  \bigskip
  \begin{center}
    {\LARGE\bf Addressing Detection Limits with Semiparametric Cumulative Probability Models}
\end{center}
  \medskip
} \fi

\bigskip
\begin{abstract}
Detection limits (DLs), where a variable is unable to be measured outside of a certain range, are common in research. Most approaches to handle DLs in the response variable implicitly make parametric assumptions on the distribution of data outside DLs. We propose a new approach to deal with DLs based on a widely used ordinal regression model, the cumulative probability model (CPM). The CPM is a type of semiparametric linear transformation model. CPMs are rank-based and can handle mixed distributions of continuous and discrete outcome variables. These features are key for analyzing data with DLs because while observations inside DLs are typically continuous, those outside DLs are censored and generally put into discrete categories.  With a single lower DL, the CPM assigns values below the DL as having the lowest rank. When there are multiple DLs, the CPM likelihood can be modified to appropriately distribute probability mass. We demonstrate the use of CPMs with simulations and two HIV data examples. The first example models a biomarker in which 15\% of observations are below a DL. The second uses multi-cohort data to model viral load, where approximately 55\% of observations are outside DLs which vary across sites and over time.

\end{abstract}

\noindent%
\vfill

\newpage
\spacingset{1.9} 
\section{Introduction}
\label{sec:intro}

Detection limits (DLs) are not uncommon in biomedical research and other fields. For example, radiation doses may only be detected above a certain threshold \citep{wing1991mortality}, antibody concentrations may not be measured below certain levels \citep{wu2001development}, and X-rays may have lower limits of detection \citep{pan2017cs}. In HIV research, viral load can only be detected above certain levels. To complicate matters, DLs often vary by assay and may change over time. For example, HIV viral load assays have had lower DLs at 400, 300, 200, 50, and 20 copies/mL depending on the commercial assay and year of application \citep{steegen2007evaluation}.

Different types of analysis methods to handle DLs have been proposed for different purposes. In this manuscript, we will focus on studying the association between an outcome variable and covariates, where the outcome variable is subject to DLs.  This is typically achieved with some sort of regression model. 
One common and simple method is to dichotomize the outcome as detectable or undetectable, and then to perform logistic regression \citep{jiamsakul2017hiv}. While it can be useful for some purposes, the dichotomization leads to information loss since the observed values inside the DLs are treated as if they are the same. 
Another common approach for handling DLs is substitution, where all nondetects are imputed with a single constant
and a linear regression model is fit. The imputed constant may be, for example, the DL itself, DL/2, DL/$\sqrt{2}$ \citep{hornung1990estimation,lubin2004epidemiologic,helsel2011statistics}, or the expectation of the measurement conditional on being outside the DL under some assumed parametric model \citep{garland1993toenail}. For example, DL/2 corresponds to the expectation of a uniform distribution between $0$ and the DL. Although simple, these substitution approaches typically result in biased estimation, underestimated variances, and thus sometimes  wrong conclusions \citep{baccarelli2005handling,fievet2010dealing}. In a third approach, one explicitly makes parametric assumptions on the distribution of the data, both within and outside the DLs. Parameters of interest can then be estimated by maximizing the censored data likelihood. Such a maximum likelihood approach is efficient and consistent when the distribution is correctly specified, but may perform poorly when distributional assumptions are incorrect. To compound the problem, there is typically no way to examine model fit outside the DLs; goodness-of-fit of a parametric model inside DLs does not ensure goodness-of-fit outside the DLs \citep{baccarelli2005handling, harel2014use}.  A related, fourth approach for addressing DLs is to multiply impute values outside the DLs \citep{little2019statistical, harel2007multiple}.  This approach may be computationally expensive, still requires parametric assumptions that can only be verified inside DLs, and may be particularly problematic with high rates of censoring or small sample sizes \citep{lubin2004epidemiologic,zhang2009nonparametric}. 

To avoid strong parametric assumptions, nonparametric methods such as Kaplan–Meier, score and rank-based methods have been proposed in two-sample comparisons \citep{helsel2011statistics}. Zhang et al. (2009) explored the use of the Wilcoxon rank sum test, other weighted rank tests, Gehan and Peto-Peto tests, and a novel nonparametric method for location-shift inference with DLs. Although attractive for two-sample tests, these nonparametric methods do not permit the inclusion of covariates.

In this manuscript, we propose a new approach for analyzing data subject to detection limits.  Data with DLs effectively follow a mixture distribution, where those below a lower DL can be thought of as belonging to a discrete category, those above an upper DL belonging to another discrete category, while those inside the DLs are continuous.  Whether discrete or continuous, the values are orderable. In earlier work, \cite{liu2017modeling} showed that continuous response variables can be modeled using a popular model for ordinal outcomes, namely the cumulative probability model (CPM), also known as the \lq cumulative link model' \citep{agresti2003categorical}. CPMs are a type of semiparametric linear transformation model, in which the continuous response variable after some unspecified monotonic transformation is assumed to follow a linear model, and the transformation is nonparametrically estimated \citep{zeng2007maximum}. These models are very flexible and can handle a wide variety of outcomes, including variables with DLs. 
Importantly, when fitting CPMs to data with DLs, minimal assumptions are made on the distribution of the response variable outside the DLs as these models are based on ranks, and values below/above DLs are simply the lowest/highest rank values.  Because of their relationship to the Wilcoxon rank sum test \citep{mccullagh1980regression}, the CPM can be thought of as a semiparametric extension to permit covariates to the approaches that \citet{zhang2009nonparametric} found effective for handling DLs in two-sample comparisons. Finally, as will be shown, because CPMs model the conditional cumulative distribution function (CDF), it is easy to extract many different measures of conditional association from a single fitted model, including conditional quantiles, conditional probabilities, odds ratios, and probabilistic indexes, which permits flexible and compatible interpretation.

In Section 2, we review the CPM, illustrate its use for simple settings where there is only a single set of DLs, and then show how CPMs can be extended to address multiple DLs. We also propose a new method for estimating the conditional quantile from a CPM. In Section 3, we illustrate and demonstrate the advantages of the proposed approach using real data from two studies.  The first study aims to measure the association between covariates and a biomarker whose values are below a DL in approximately 15\% of observations. The second example is a large multi-cohort study of viral load (VL) after starting antiretroviral therapy among persons with HIV, where most observations are below DLs, but the DLs vary across sites and change over time.  In Section 4, we demonstrate the performance of our method with simulations. The final section contains a discussion of the strengths and limitations of our method and future work.

\section{Methods}
\label{sec:meth}

\subsection{Cumulative Probability Models}

Transformation is often needed for regression of a continuous outcome
variable $Y$ to satisfy model assumptions, but specifying the correct
transformation can be difficult. In a linear transformation model, the
outcome is modeled as $Y=H(\beta^T X+\epsilon)$, where $H(\cdot)$ is an
unknown monotonically increasing transformation, $X$ is a vector of
covariates, and $\epsilon$ follows a known distribution with CDF
$F_\epsilon$. This linear transformation model can be equivalently expressed
in terms of the conditional CDF,
\begin{align*}
  F(y|X) \equiv \Pr(Y\le y |X) 
  =\Pr[\epsilon\le H^{-1}(y)-\beta^T X|X]
  =F_\epsilon[H^{-1}(y)-\beta^T X].
\end{align*}
Let $G=F_\epsilon^{-1}$ and $\alpha=H^{-1}$; $\alpha(\cdot)$ is monotonically
increasing but otherwise unknown.  Then
\begin{equation}
  G[F(y|X)]=\alpha(y)-\beta^T X,
  \label{eq:CPM}
\end{equation}
where $G$ serves as a link function and the model becomes a cumulative
probability model (CPM). The intercept function $\alpha(y)$ is the transformation of the response variable such that $\alpha(Y)=\beta^TX + \epsilon.$ 
The $\beta$ coefficients indicate the association between the response variable and covariates: fixing other covariates, a positive/negative $\beta_j$ means that an increase in $X_j$ is associated with a stochastic increase/decrease in the distribution of the response variable.

In the CPM (\ref{eq:CPM}), the intercept function
$\alpha(y)$ can be nonparametrically estimated with a step function
 \citep{zeng2007maximum, liu2017modeling}. This allows great model flexibility. 
 Consider an iid dataset
$\{(y_i,x_i): i=1,\ldots, n\}$.  The nonparametric likelihood is
\begin{equation}
  \prod_{i=1}^n \left[F (y_i|x_i)-F(y_i^-|x_i)\right],
  \label{eq:nplikelihood}
\end{equation}
where $F(y_i^-|x_i)=\lim_{t \uparrow y_i}F(t|x_i)$.  In nonparametric maximum likelihood estimation, the probability mass given any $x$ will be
distributed over the discrete set of observed outcome values. Thus we only need to consider functions for $\alpha(\cdot)$
such that $F(y|x_i)$ is a discrete distribution over the observed values.
Let $J$ be the number of distinct outcome values, denoted as
$a_{1}<\cdots<a_{J}$.  Let $S=\{a_1,\ldots,a_J\}$. These serve as the anchor points for the nonparametric
likelihood.  Let $\alpha_j =\alpha(a_{j})$; then $\alpha_1<\cdots<\alpha_J$.
The nonparametric likelihood (\ref{eq:nplikelihood}) can be written as
\begin{align}
  L(\beta, \bm{\alpha})=&\prod_{i:y_i=a_1} F_{\epsilon}(\alpha_1-\beta^T x_i) \times
    \prod_{j=2}^{J-1}\prod_{i:y_i=a_j}\left[F_{\epsilon}(\alpha_j-\beta^T
    x_i)-F_{\epsilon}(\alpha_{j-1}-\beta^Tx_i)\right] \times
    \\
  & \prod_{i:y_i=a_{J}}\left[1-F_{\epsilon}(\alpha_{J-1}-\beta^Tx_i)\right].
    \label{eq:final_l}
\end{align}
Maximizing (\ref{eq:final_l}), we obtain the nonparametric maximum likelihood estimates (NPMLEs), $(\hat{\beta}, \bm{\hat{\alpha}})$, where
$\bm{\hat\alpha} =(\hat \alpha_1, \dots, \hat \alpha_{J-1})$.  
Note the multinomial form of the likelihood (\ref{eq:final_l}); because the probabilities in a multinomial likelihood add to one, $\alpha_J$ is not estimated.   
Note also that the likelihood in (\ref{eq:final_l}) is
identical to that of cumulative link models for ordinal data if the outcome
$Y$ is treated as ordinal with categories $\{a_1,\ldots,a_J\}$.
\citet{liu2017modeling} and \citet{tian2020empirical} have shown that CPMs can be fit to and work well
for continuous and mixed types of responses. CPMs have also been shown to be
consistent and asymptotically normal, with variance consistently estimated
with the inverse of the information matrix under mild conditions including
boundedness of the outcome variable \citep{li2022asymptotics}. The NPMLEs and their estimated variances can be
efficiently computed with the \texttt{orm()} function in the \textbf{rms}
package in \textsf{R} \citep{rms}, which takes advantage of the tridiagonal nature of the Hessian matrix using Cholesky decomposition \citep{liu2017modeling}.  

CPMs have several nice features. Some widely used regression methods model only one aspect of the conditional distributions (e.g., conditional mean for linear regression and conditional quantile for quantile regression). With the NPMLEs
$(\hat{\beta}, \bm{\hat{\alpha}})$, we can estimate the conditional CDFs as
$\hat{F}(y|x)=F_\epsilon(\hat{\alpha}_j - \hat{\beta}^T x)$ where $j$ is the index such that $a_j=\max\{a \in S: a \leq y\}$;
standard errors can be obtained by the delta method.  Since conditional
CDFs are directly modeled, other characteristics of the distribution, such as
the conditional quantiles and conditional expectations, can be easily derived
\citep{liu2017modeling}.  Depending on the choice of link function, $\beta$ may be
interpretable; for example, with the logit link function, exp$(\beta)$ is an
odds ratio.  Probabilistic indexes \citep{de2019semiparametric}, which are defined as $\Pr(Y_1<Y_2|X_1,X_2)$,
can also be easily derived; for example, with the logit link, $P(Y_1 < Y_2|X_1, X_2)=\left[1+\exp\left(-(X_2-X_1)^T \beta \right)\right]^{-1}$.
With the transformation $\alpha(\cdot)$
nonparametrically estimated, CPMs are invariant to any monotonic
transformation of the outcome; therefore, no pre-transformation is needed. With a single binary covariate and the logit link function, the score test for the CPM is nearly identical to the Wilcoxon rank sum test \citep{mccullagh1980regression}; see Supplemental Material S1.1. Because only the order of the outcome values
but not the specific values matter when estimating $\beta$ in the CPM, it can handle any ordinal,
continuous, or mixture of ordinal and continuous distributions, which can be
useful for analyzing data with DLs. 

\subsection{Single Detection Limits}

In this subsection, we first present our method for the simple scenario that there is a single lower DL and/or a single upper DL.  We will describe the general approach for multiple DLs in the next subsection.

Consider a dataset with a lower DL, $l$, and an upper DL, $u$.  The outcome $Y$ is
observed if it is inside the DLs (i.e., $l\le Y\le u$) or censored if it is
outside the DLs.  The $J$ distinct values of the observed outcomes are denoted as $l\le a_{1}<\cdots<a_{J}\le u$.  When there are no observations outside the DLs, these values are treated as ordered categories in CPMs and they are the anchor points in the nonparametric likelihood (\ref{eq:final_l}), and
correspondingly there are $J-1$ alpha parameters, $\alpha_1<\cdots<\alpha_{J-1}$.
With observations outside the DLs, the likelihood (\ref{eq:final_l}) needs to be modified accordingly.

When there are observations below the lower DL, we do not know their values except that they are $<l$.  As there is no way to distinguish them, we
treat them as a single category, denoted as $a_{0}$.  Note that
$a_{0}$ is not a value but a symbol for the additional category below
$a_{1}$.  The nonparametric likelihood for a subject outcome censored at the lower DL
$l$ is
\begin{equation*}
  \Pr(Y_i<l|X_i=x_i) = F_\epsilon(\alpha_0-\beta^Tx_i),
\end{equation*}
where $\alpha_0$ is the extra alpha parameter corresponding to category
$a_{0}$ such that $\alpha_0<\alpha_1$.  Because $a_{1}$, the previously
lowest category, now has a category below it, the nonparametric likelihood
for a subject with $y_i=a_{1}$ becomes
\begin{equation*}
  F_\epsilon(\alpha_1-\beta^Tx_i) - F_\epsilon(\alpha_0-\beta^Tx_i).
\end{equation*}

Similarly, when there are observations above the upper DL, they are also
treated as a single category, denoted as $a_{J+1}$, which is a symbol for the
additional category above $a_{J}$.  The nonparametric likelihood for a
subject censored at the upper DL $u$ is
\begin{equation*}
  \Pr(Y_i>u|X_i=x_i) = 1-F_\epsilon(\alpha_J-\beta^Tx_i),
\end{equation*}
Because $a_{J}$ is no longer the highest
category, $\alpha_{J}$ will need to be estimated, and the likelihood for a
subject with $y_i=a_{J}$ is now
\begin{equation*}
  F_\epsilon(\alpha_J-\beta^Tx_i) -F_\epsilon(\alpha_{J-1}-\beta^Tx_i).
\end{equation*}

Put together, with observed data subject to a single lower DL and a single upper DL, the CPM likelihood is 
\begin{align}
  L(\beta, \bm{\alpha})=&\prod_{i:y_i=a_0} F_{\epsilon}(\alpha_0-\beta^T x_i) \times
    \prod_{j=1}^{J}\prod_{i:y_i=a_j}\left[F_{\epsilon}(\alpha_j-\beta^T
    x_i)-F_{\epsilon}(\alpha_{j-1}-\beta^T x_i)\right] \times \nonumber\\
  &\quad \prod_{i:y_i=a_{J+1}}\left[1-F_{\epsilon}(\alpha_{J}-\beta^T x_i)\right],
\label{eq:final_singleDL}
\end{align}
which is equivalent to (\ref{eq:final_l}) except with two new anchor points, $a_0$ and $a_{J+1}$. Therefore, (\ref{eq:final_singleDL}) is maximized in an identical manner to (\ref{eq:final_l}), with outcomes below the lower DL and outcomes above the upper DL simply assigned to categories $a_0$ and $a_{J+1}$, respectively. 

In summary, when there are data censored below
the lower DL, we add a new anchor point $a_{0}<a_{1}$ and a new parameter
$\alpha_0$; when there are data censored above the upper DL, we add a new
anchor point $a_{J+1}>a_{J}$ and a new parameter $\alpha_{J}$.  The alpha parameters to be estimated are
$(\alpha_1,\ldots,\alpha_{J-1})$ when there are no DLs or no data censored at
DLs, $(\alpha_0,\alpha_1,\ldots,\alpha_{J-1},\alpha_J)$ when both categories
$a_{0}$ and $a_{J+1}$ are added, $(\alpha_0,\alpha_1,\ldots,\alpha_{J-1})$
when only $a_{0}$ is added, and $(\alpha_1,\ldots,\alpha_{J-1},\alpha_J)$
when only $a_{J+1}$ is added.

In practice, one can fit the NPMLE in these settings using the \texttt{orm()} function by setting outcomes below the lower DL to some arbitrary number $<l$ and outcomes above the upper DL to some arbitrary number $>u$.  Note that unlike single imputation approaches for dealing with DLs, the CPM estimation procedure is invariant to the choice of these numbers assigned to values outside the DLs. The CPM (\ref{eq:CPM}) assumes that after some unspecified transformation, the outcome follows a linear model both within and outside the DLs. In contrast, parametric approaches to deal with DLs assume the full distribution of the outcome conditional on covariates is known, both within and outside DLs. Hence, CPMs make much weaker assumptions than fully parametric approaches. 

\subsection{Multiple Detection Limits}

We now consider the general situation where data may be collected from
multiple study sites.  A site may have no DL, only one DL, or both lower and
upper DLs.  Each site may have different lower DLs and different upper
DLs, which may change over time.

Every subject has a vector $X$ of covariates and three underlying random
variables $(Y, C_L, C_U)$, where $Y$ is the true outcome and $C_L<C_U$ are the
lower and upper DLs.  When there is no upper DL, $C_U=\infty$, and when there is no lower DL, $C_L=-\infty$. $C_L$ and $C_U$ are assumed to be independent of $Y$ conditional on $X$; the vector $X$ may contain variables for study sites or calendar time. This non-informative censoring assumption is typically plausible as DLs are determined by available equipment/assays. 

We assume the CPM  (\ref{eq:CPM}) holds for all subjects.
Due to DLs, we may not always observe $Y$.  Instead, we only observe
$(Z,\Delta)$, where $Z=\max(\min(Y,C_U),C_L)$ and $\Delta$ is a variable
indicating whether $Y$ is observed or censored at a DL: $\Delta=1$ and $Z=Y$ if
$Y$ is observed, $\Delta=L$ and $Z=C_L$ if $Y<C_L$, and $\Delta=U$ and
$Z=C_U$ if $Y>C_U$.  

Given a dataset $\{(z_i,\delta_i;x_i)\}$ ($i=1,\ldots,n$), we first determine how many anchor points are needed to support the nonparametric likelihood of the CPM.  Let $J$ be the number of distinct values of $z_i$ among those with $\delta_i=1$; they are denoted as $a_{1}<\cdots<a_{J}$.  For data without any
DLs, these points are the anchor points for the nonparametric likelihood, and
they are effectively treated as ordered categories in a CPM.  Let
$S=\{a_{1},\cdots,a_{J}\}$ be the set of these values.  When there are
data with $\delta_i=L$, let $l$ be the smallest $z_i$ with $\delta_i=L$.
Similarly, when there are data with $\delta_i=U$, let $u$ be the largest
$z_i$ with $\delta_i=U$.  If $l\le a_{1}$, we add a category into $S$ below
$a_{1}$, denoted as $a_{0}$; note that it is not a value but a symbol for the
additional category in $S$ below $a_{1}$.  Similarly, if $u\ge a_{J}$, we add
$a_{J+1}$ into $S$, which is a symbol for the additional category above
$a_{J}$.  Depending on the data, the number of ordered categories can be $J$,
$J+1$, or $J+2$.

Consider the situation where both $a_{0}$ and $a_{J+1}$ have been added
to $S$ (i.e., $S=\{a_{0},a_{1},\ldots,a_{J},a_{J+1}\}$).  When
$\delta_i=1$, the nonparametric likelihood for $(z_i,1)$ is
\begin{equation}
  F_\epsilon(\alpha_{j}-\beta^Tx_i) - F_\epsilon(\alpha_{j-1}-\beta^Tx_i),
\end{equation}
where $j$ is the index such that $a_{j}=z_i$.  When $\delta_i=L$, the
nonparametric likelihood for $(z_i,L)$ is
\begin{equation}
  \Pr(Y<z_i|x_i)= \left\{
  \begin{array}{lr}
    F_\epsilon(\alpha_{0}-\beta^Tx_i), & (z_i= l) \\
    F_\epsilon(\alpha_{j}-\beta^Tx_i), & (z_i\ne l)
  \end{array} \right.
  \label{eq:likelihoodL}
\end{equation}
where $j$ is the index such that $a_{j}=\max\{a\in S:a<z_i\}$ when
$z_i\ne l$.  When $\delta_i=U$, the nonparametric likelihood for $(z_i,U)$ is
\begin{equation}
  \Pr(Y>z_i|x_i)= \left\{
  \begin{array}{lr}
    1-F_\epsilon(\alpha_{J}-\beta^Tx_i), & (z_i= u) \\
    1-F_\epsilon(\alpha_{j-1}-\beta^Tx_i), & (z_i\ne u)
  \end{array} \right.
  \label{eq:likelihoodU}
\end{equation}
where $j$ is the index such that $a_{j}=\min\{a\in S:a>z_i\}$ when
$z_i\ne u$.  The overall nonparametric likelihood is the product of these
individual likelihoods over all subjects. Note that if there are no uncensored observations between two lower (or upper) DLs, the two DLs are effectively treated as the same DL. A toy example to illustrate our definition is provided in Table S1 of the Supplementary Material.

Slight modifications will be applied when no or only one additional category
is added to $S$.  When there is no need to add $a_{0}$ to $S$ (i.e., when $l>a_1$ or there are no lower DLs), only the second
row in the likelihood (\ref{eq:likelihoodL}) for $(z_i,L)$ will be employed, and the likelihood
for $(z_i,1)$ with $z_i=a_{1}$ is $F_\epsilon(\alpha_{1}-\beta^Tx_i)$.  When
there is no need to add $a_{J+1}$ to $S$ (i.e., when $u<a_J$ or there are no upper DLs), only the second row in the
likelihood (\ref{eq:likelihoodU}) for $(z_i,U)$ will be employed, and the likelihood
for $(z_i,1)$ with $z_i=a_{J}$ is $1-F_\epsilon(\alpha_{J-1}-\beta^Tx_i)$.

Similar to the likelihood of CPM for data without DLs, the individual
likelihoods presented above involve either one alpha parameter or two adjacent
alpha parameters.  As a result, the Hessian matrix continues to be
tridiagonal, allowing us to use Cholesky decomposition to solve for the NPMLEs and efficiently estimate their variances.
We have developed an \textsf{R} package, \textbf{multipleDL} available at https://github.com/YuqiTian35/multipleDL, which uses the \texttt{optimizing()} function in the \textbf{rstan} package to maximize the likelihood \citep{rstan}. 

\subsection{Interpretable Quantities and Conditional Quantiles}

Interpretation of results after fitting CPMs to outcomes with DLs is similar to settings without DLs. Depending on the link function, $\beta$ may be directly interpretable.  The conditional CDF, probabilistic indexes, and conditional quantiles are also easily derived. 
Note, however, that without additional assumptions on the distribution of the outcome outside DLs, conditional expectations cannot be estimated. 

\begin{figure}
    \centering
    \includegraphics[scale=0.7]{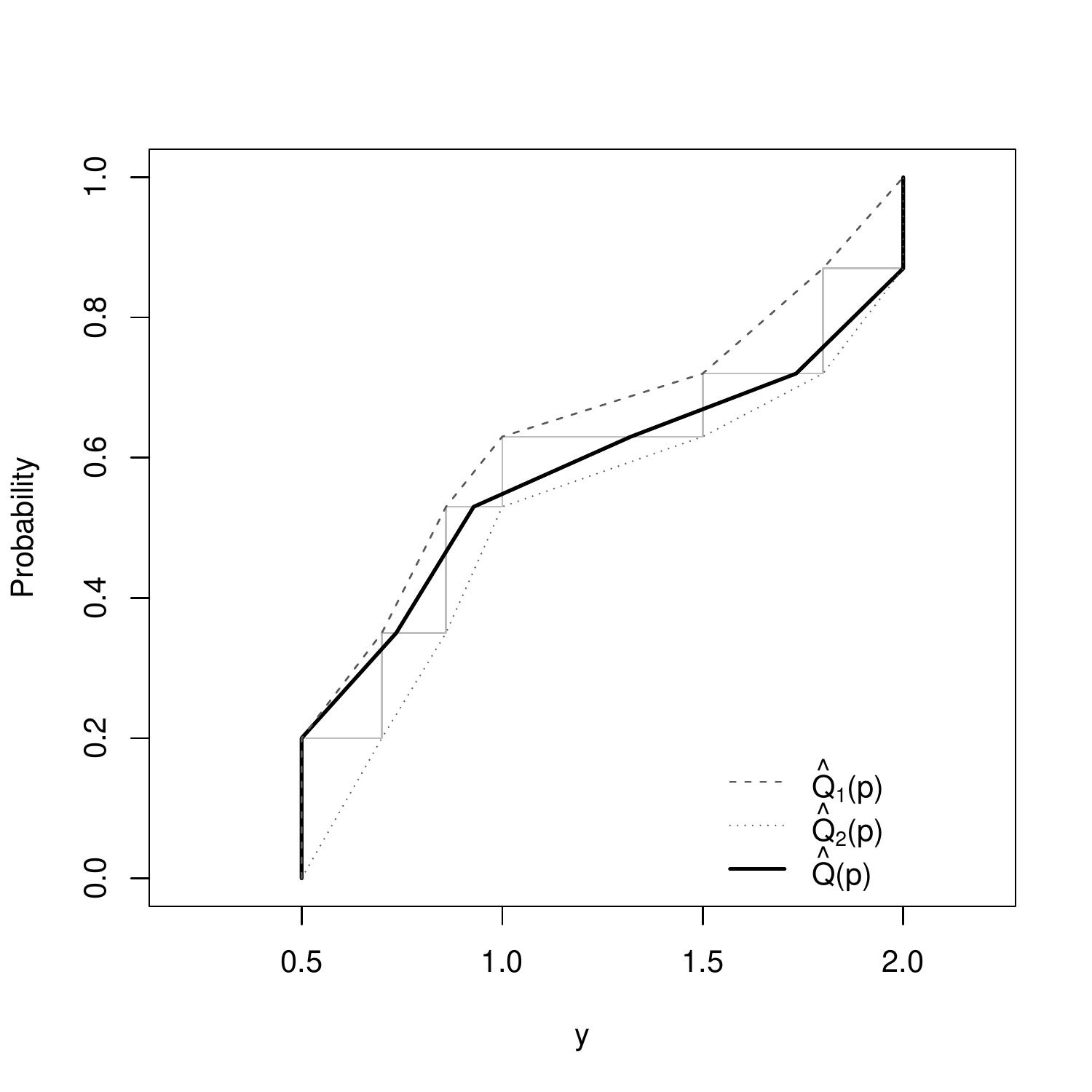}
\caption{Illustration of three approaches for conditional quantiles. The data set has a lower DL 0.5, an upper DL 2, and five observed values of $y$: 0.7, 0.86, 1, 1.5, 1.8. Thus $S=\{\text{`}{<}0.5\text{'}, 0.7, 0.86, 1, 1.5, 1.8, \text{`}{>}2\text{'} \}$. The dashed lines are for $\hat{Q}_1(p)$, the dotted lines are for $\hat{Q}_2(p)$, the solid black lines are for $\hat{Q}(p)$, and the solid gray lines are for the empirical CDF.  Here, $\hat Q(p) =\hat Q_1(p)= \text{`}{<}0.5\text{'}$ when $p<\hat F(0.5|x)$, and $\hat Q(p) =\hat Q_2(p)=\text{`}{>}2\text{'}$ when $p>\hat F(2|x)$.}
    \label{fig:quantile}
\end{figure}

We now describe how to infer conditional quantiles from a CPM fitted on data
with DLs.  The conditional CDF from a CPM for a given $x$ can be computed
as $\hat F(y|x) =F_\epsilon(\hat\alpha_j-\hat\beta^Tx)$ where $j$ is the index such that $a_{j}=\max\{a\in S:a\le y\}$; if there is no $a\in S$ such that $a\le y$, then $\hat F(y|x)=0$.  For ease of presentation,
we fix $x$ and let $P_j=\hat F(a_j|x)$ ($j=0,1,\ldots,J,J+1$); for
convenience, let $P_{-1}=0$.  Our goal is to define a quantile function
$\hat Q(p)$, where $0<p<1$, for the conditional distribution given $x$.

The quantile function for a CDF $F(\cdot)$ is typically defined as
$Q(p)=\inf\{z:F(z) \ge p\}$.  A plug-in estimator for an estimated CDF,
$\hat F$, is $\hat{Q}_0(p)=\inf\{z:\hat{F}(z) \ge p\}$.  When applied
to our setting, $\hat Q_0(p)=a_j$ when $P_{j-1}<p\le P_j$.  This estimator may
not be suitable for CPMs because $\hat F(\cdot)$ is a step function and therefore $\hat Q_0(p)$ only takes values at the anchor points,
which can be undesirable for continuous outcomes, especially when there is a
large gap between adjacent anchor points.  

\citet{liu2017modeling} proposed to estimate quantiles for CPMs with linear
interpolation. Specifically, given a fixed $p$, let $j=j(p)$ be the index
such that $P_{j-1}<p\le P_j$.  When $p>P_0$, $j\ge 1$ and define
$\hat Q_1(p) =a_{j-1} + (p-P_{j-1})/ (P_{j}-P_{j-1})\times (a_{j} - a_{j-1})$,
which is a linear interpolation between $a_{j-1}$ and $a_{j}$.  When
$p\le P_0$, $\hat Q_1(p)$ is set to be $a_0$. Recall that $a_0$ is not a value but a symbol for being below the lower DL, $l$; we thus relabel it as `${<}l$', so when $p\le P_0$, $\hat Q_1(p)=$ `${<}l$'. For the linear interpolation between $a_0$ and $a_1$, we set $a_0$ to be $l$. Similarly, $a_{J+1}$ is labeled `${>}u$' and assigned the value $u$ for the linear interpolation between $a_J$ and $a_{J+1}$. $\hat Q_1(p)$ is illustrated
as the dashed lines in Figure \ref{fig:quantile}.  An alternative definition is
to interpolate between $a_{j}$ and $a_{j+1}$:
$\hat Q_2(p)=a_{j}+ (p- P_{j-1}) / (P_{j}-P_{j-1}) \times (a_{j+1} - a_{j})$ when
$p<P_J$ and $\hat Q_2(p)=a_{J+1}=$ `${>}u$' when $p\ge P_J$.  $\hat Q_2(p)$ is illustrated as the
dotted lines in Figure \ref{fig:quantile}.  For continuous data without DLs,
$\hat Q_1(p)$ and $\hat Q_2(p)$ converge as the sample size increases.
However, they are problematic for continuous data with DLs because
$\hat Q_1(p)<a_{J+1}$ for all $p<1$ and $\hat Q_2(p)>a_{0}$ for all $p>0$
even though there are non-zero estimated probabilities at the lower DL $a_0$
and upper DL $a_{J+1}$.

We propose a new quantile estimator as a weighted average between
$\hat Q_1(p)$ and $\hat Q_2(p)$, 
\begin{equation}
  \hat Q(p)=(1-w) \hat Q_1(p) + w \hat Q_2(p),
\end{equation}
where $w=w(p) = (p-P_{0}) / (P_{J}-P_{0})$ when $P_0<p<P_J$, $0$ when
$p\le P_0$, and $1$ when $p\ge P_J$.  This definition is shown as the black
curve in Figure \ref{fig:quantile}.
Note that $\hat Q(p)$ equals $\hat Q_1(p)=$ `${<}l$' when $p\le P_0$, and equals
$\hat Q_2(p)=$ `${>}u$' when $p\ge P_J$.  It can be shown that similar to $\hat Q_1(p)$ and $\hat Q_2(p)$, $\hat Q(p)$ is also piecewise linear with transition points at $P_j$ ($j=0,1,\ldots,J$).

In situations where there is only a lower DL or an upper DL, our definition of $\hat Q(p)$ is similar. Confidence intervals for the conditional quantiles can be estimated by applying a weighted linear interpolation to the confidence intervals of the conditional CDF similar to the above procedure \citep{liu2017modeling}.

\section{Applications}

In this section, we illustrate our method with two datasets, one from a biomarker study with a single lower DL and the other from a multi-center study with multiple DLs varying within and across centers.  

\subsection{Single Detection Limit}
Our first example uses data from a study investigating the relationship between HIV, diabetes, obesity, and various biomarkers.  
Data were collected on 161 adults, some of whom were highly overweight (body mass index (BMI) ranged from 22 to 58 kg/m\textsuperscript{2}).  Several biomarkers were measured.  Here, we focus on interleuken 4 (IL-4), a cytokine that is related to T-cell production and metabolism and has been seen to limit lipid accumulation in mice \citep{tsao2014interleukin}.  We examine the association between IL-4 and BMI, controlling for age, sex, HIV status, and diabetes status.  Our measures of IL-4 had a single lower DL of 0.019 pg/ml, and 24 subjects (15\%) had IL-4 values below the DL. The distribution of IL-4, which is right-skewed, is shown in Figure S1(A) in Supplemental Material S2.1.

\begin{figure}
    \centering
    \includegraphics[scale=0.75]{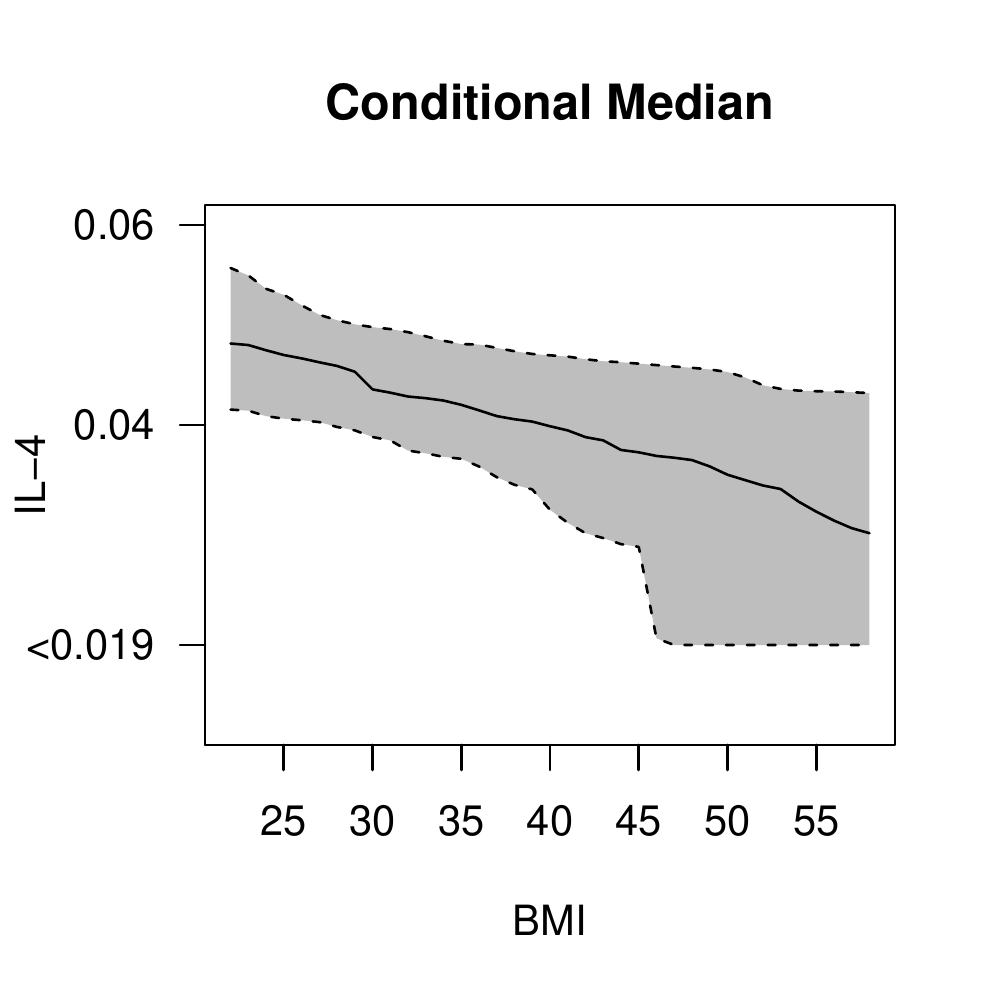}
    \caption{The conditional median obtained by CPMs varying BMI while fixing other covariates at median/mode levels}
    \label{fig:median_cpm}
\end{figure}

We fit a CPM as described in Section 2.2, using the logit link; full results are in Table S2 in the Supplementary Material S2.1.  No transformation of IL-4 was needed.  With the logit link function, the $\beta$ parameters can be interpreted as log odds ratios. BMI was found to be negatively associated with IL-4 (p-value 0.023). Holding other covariates constant, a 5 kg/m$^2$ increase in BMI corresponded to a 22\% decrease in the odds of having a higher IL-4 value (adjusted odds ratio 0.78, 95\% confidence interval (CI) of $(0.62, 0.97)$).  The corresponding probabilistic index was 0.264, meaning that holding other variables constant, a 5 kg/m$^2$ increase in BMI was associated with a 0.736 ($=1-0.264$) probability of having a lower IL-4.  
The median IL-4 conditional on BMI and controlling for all other covariates at their median/mode levels were estimated from the CPM and is shown in Figure \ref{fig:median_cpm}. The conditional median decreased as BMI increased, with the 95\% CI including the category `$<$0.019' for those with a very large BMI. Note that `$<$0.019' is the smallest ordered category indicating values below the DL. Other quantiles and quantities can also be easily derived from the CPM; for example, Figure S2 in the Supplementary Material shows the 90th percentile of IL-4 as a function of BMI, and the probabilities of IL-4 being greater than 0.019 (the DL) and greater than 0.05 as functions of BMI.

It is worth comparing results from the CPM to other potential analysis approaches. (i) The most common approach in practice would be to singly impute those values below the DL; given the skewed nature of the data, one would then likely log-transform the data and fit a linear regression model.  The result can vary depending on the choice of the imputed number: if one imputes with the DL itself (0.019) vs.\ $0.001$, the log-transformed IL-4 is estimated to decrease $0.013$ pg/ml vs.\ $0.032$ pg/ml, respectively, per 5 kg/m$^2$ increase in BMI, with different statistical significance (p-value 0.020 vs.\ 0.073).  (ii) A more sophisticated approach might be to assume the data are log-normally distributed and perform a likelihood-based analysis, which results in an estimated change on the log-scale of $-0.015$ pg/ml per 5 kg/m$^2$ BMI increase (p-value 0.018).
The conditional median as a function of BMI could also be extracted from this analysis and is in Figure S3(B).  The curve of conditional median as a function of BMI is similar to what was estimated with the CPM (Figure S3(A)), but it is slightly lower and its confidence bands are tighter than those of the CPM.  The tighter bands reflect the parametric assumption that the data are truly log-normally distributed.  In contrast, the CPM does not require transformation of the data, and it non-parametrically estimates the best transformation.  (iii) One could also directly estimate the conditional median as a function of BMI using quantile regression \citep{koenker2001quantile}. This estimated curve is in Figure S3(C), which closely matches that estimated from the CPM. However, the confidence bands for median regression are wider than those of the CPM, and the 95\% CI for the slope contains 0. One could argue that the CPM is assuming more than median regression (which only assumes a linear relationship between the median on the original outcome scale and the covariates); hence the narrower confidence bands.  However, the CPM is able to yield several additional quantities (e.g., other quantiles, odds ratios, exceedance probabilities) from a single model that cannot be obtained from median regression. Also, the confidence bands obtained by the CPM do not go below the DL. (iv) Finally, one could dichotomize IL-4 into ``undetectable'' and ``detectable'' and fit a logistic regression model.  However, logistic regression was not able to provide a stable estimation for this dichotomization. One could consider other dichotomizations, but the choice is arbitrary.  In fact, a beta coefficient in the CPM can be thought of as a weighted average of the log-odds ratios for logistic regression models that consider all possible orderable dichotomizations of the outcome. 

\subsection{Multiple Detection Limits}
\begin{figure}
    \centering
    \includegraphics[scale=0.55]{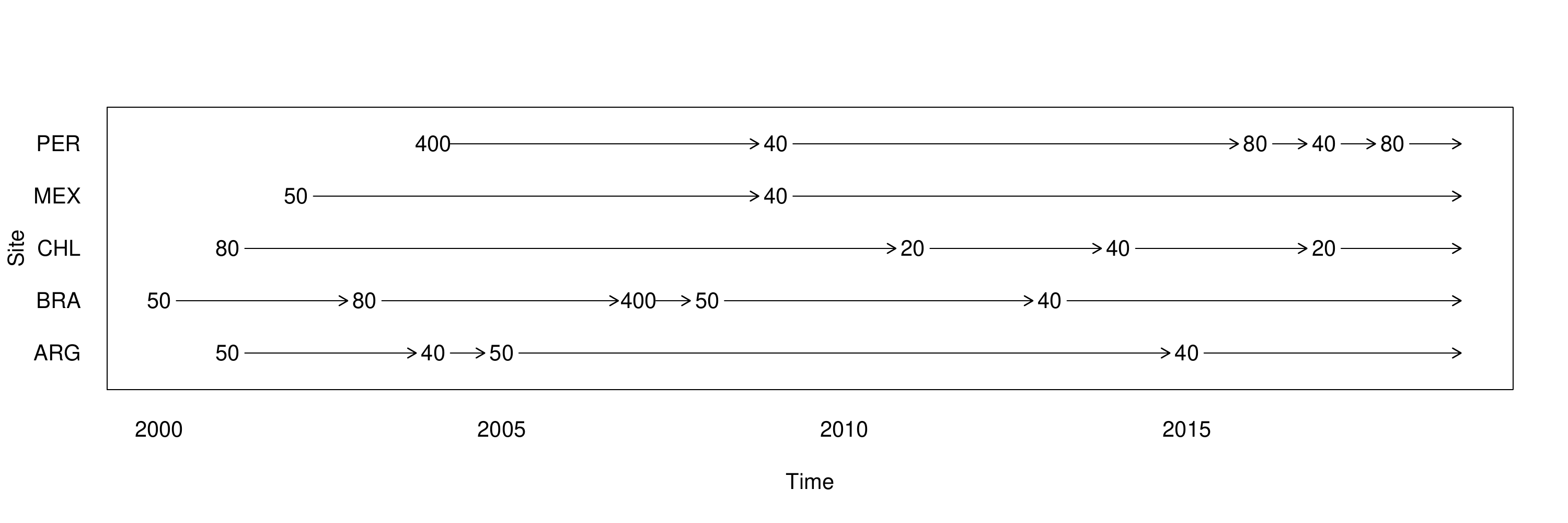}
    \caption{The changes of most frequent DL values every year at each study site over time.}
    \label{fig:dl_change}
\end{figure}

We illustrate our approach to handle multiple DLs with data from a multi-center HIV study. The data include 5301 adults living with HIV starting antiretroviral therapy (ART) at one of 5 study centers in Latin America 
between 2000 and 2018. Viral load (VL) measures the amount of virus circulating in a person with HIV. A high VL after ART initiation may indicate non-adherence or an ineffective ART regimen that should be switched. We study the association between VL at approximately 6 months after ART initiation and variables measured at ART initiation (baseline).
The DLs for the outcome VL differed by site and calendar time. Figure \ref{fig:dl_change} shows the most frequent lower DL values for each year and at each site.  There are five distinct lower DLs in this database: 20, 40, 50, 80, and 400 copies/mL. 
A total of 2992 (56\%) patients had 6-month VL censored at a DL: 45\%, 54\%, 52\%, 65\%, and 57\% at study sites in Argentina, Brazil, Chile, Mexico, and Peru, respectively. More study details are in Supplemental Material S2.2.

A traditional analysis in the HIV literature would dichotomize VL as detectable and undetectable and perform logistic regression \citep{jiamsakul2017hiv}.  There are a few issues that make this analysis less than ideal. First, all VLs above the DL (nearly half of all observations) would be collapsed into a \lq\lq detectable" category resulting in well-known loss of information due to dichotomizing continuous variables \citep{fedorov2009}. Second, because the DL varies with time and by site, the analyst is forced to dichotomize at the largest DL (in this case 400 copies/mL) or else perform an analysis where values above the DL at one site are treated differently than they would be treated at another site. For example, a VL of 300 copies/mL measured in Mexico in 2005 would be measured as `${<}400$' that same year in Peru; assigning this value as `${<}400$' results in lost information but leaving it as \lq\lq detectable" would make the outcome variable different across time and sites. A more parametric analysis might assume that the VL follows a specified distribution (e.g., log-normal distribution) and fit the censored data likelihood or multiply impute values below the DL from the assumed distribution to obtain estimated regression coefficients. However, distributional assumptions for values below the DL are strong and untestable, and given that over half of the response variables are below the DL, these assumptions would have a large impact on results.

In contrast, the CPM uses all available information (i.e., does not dichotomize the response variable) and makes much weaker assumptions than the fully parametric approaches. Similar to the parametric approaches, the CPM assumes non-informative censoring conditional on covariates (which is reasonable, given that DLs are determined by equipment/assays independent of true values) and that all observations follow a common distribution conditional on covariates, which permits borrowing information across sites and time. Unlike the fully parametric approach, however, the CPM does not fully specify this distribution. Rather, the CPM assumes that response variables follow a linear model with known error distribution after some unspecified transformation. 

\begin{table}[]
    \centering
    \caption{The $\beta$ coefficients in CPMs can be interpreted as log odds ratios. We show the odds ratio (95\% confidence interval) and p-value for the predictors included in the model.}
    \label{tab:odds_ratio}
    \renewcommand{\arraystretch}{0.53} 
    \begin{tabular}{lcc}
        \hline
        Predictor & Odds Ratio (95\% CI) & P-value \\
        \hline
        \textbf{Age} (per 10 years) &	0.98 (0.93, 1.03)	& 0.418 \\

        \textbf{Sex} & & 0.201 \\
        \hspace{3mm} Male (reference) & 1 & \\
        \hspace{3mm} Female & 0.90	(0.76, 1.06) & \\
        
        \textbf{Study center} & & ${<}0.001$ \\
        \hspace{3mm} Peru (reference) & 1 & \\
        \hspace{3mm} Argentina  & 1.26 (0.98, 1.61) & \\
        \hspace{3mm} Brazil  &	1.07 (0.91,	1.26) & \\
        \hspace{3mm} Chile  &	1.07 (0.90,	1.26) & \\
        \hspace{3mm} Mexico  &	0.59 (0.49,	0.70) & \\
        
        \textbf{Route of infection} &  & ${<}0.001$ \\
        \hspace{3mm} Homosexual/Bisexual (reference) & 1 & \\
        \hspace{3mm} Heterosexual & 0.96 (0.83, 1.10) & \\
        \hspace{3mm} Other/Unknown & 0.79 (0.62, 1.01) & \\
        
        \textbf{Prior AIDS event} & & 0.001 \\
        \hspace{3mm} No (reference) & 1 & \\
        \hspace{3mm} Yes & 1.24	(1.09,	1.41) & \\
        
        \textbf{Baseline CD4} (per 1 square root cells/$\mu$L) & 1.09	(1.08,	1.10) & 	${<}0.001$ \\
        
        \textbf{Baseline VL} (per 1 $\log_{10}$ copies/mL) &	1.44	(1.34,	1.54) & ${<}0.001$ \\
        
        \textbf{ART regimen} & & 0.034 \\
        \hspace{3mm} NNRTI-based (reference) &	1 & \\
        \hspace{3mm} INSTI-based & 0.55 (0.40, 0.75) & \\
        \hspace{3mm} PI-based & 1.10 (0.95, 1.29) & \\
        \hspace{3mm} Other & 2.57 (1.28, 5.16) & \\
        
        \textbf{Months to VL measure} & 0.95	(0.92, 0.98) & 	0.002 \\
        
        \textbf{Calendar year} & 0.89	(0.88,	0.91) & 	${<}0.001$\\
        \hline
    \end{tabular}
\end{table}

\begin{figure}
    \centering
    \includegraphics[scale=0.7]{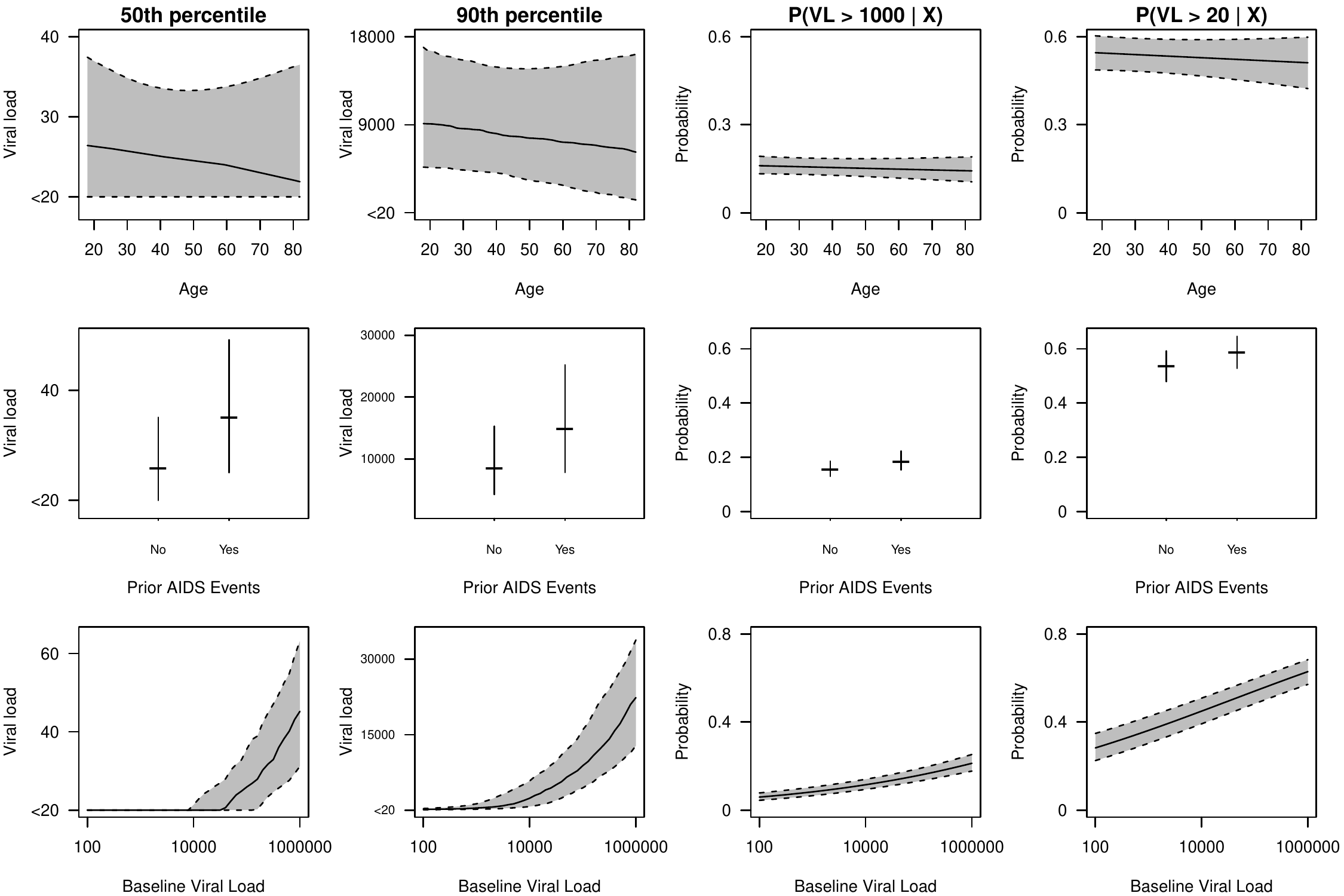}
    \caption{The estimated conditional 50th and 90th percentiles of 6-month VL and the conditional probability of 6-month VL being greater than 1000 and 20 as functions of age (top row), prior AIDS events (middle row), and baseline VL (bottom row) while keeping other covariates at their medians (for continuous variables) or modes (for categorical variables) based on our method.}
    \label{fig:age_aids_vl}
\end{figure}

We applied our method in Section 2.3 to fit a CPM of the 6-month VL on baseline variables with the logit link. Results 
are shown in Table \ref{tab:odds_ratio}. With the logit link, the $\beta$ parameters can be interpreted as log odds ratios and are presented as odds ratios in the table along with 95\% CIs. P-values are likelihood ratio test p-values. 
The results suggest that study center, route of infection,  prior AIDS event, baseline CD4 count, baseline VL, ART regimen, the time from ART initiation until the VL measurement, and calendar year are all associated with VL at 6 months. Holding other variables fixed, a 10-fold increase in VL at baseline is associated with a 44\% increase in the odds of having a higher VL at 6 months (95\% CI 34\% to 54\%).

Quantiles and cumulative probabilities are also easily extracted from the CPM. The first row of Figure \ref{fig:age_aids_vl} are the estimated conditional 50th and 90th percentiles of 6-month VL and the conditional probabilities for 6-month VL being greater than 1000 and 20 copies/mL as functions of age. The plots show that VL at 6 months is fairly similar across ages after fixing the other covariates. The smallest DL is 20 copies/mL, and all VL less than this DL belong to the smallest ordered category, which we label as `${<}20$'.
The second row of Figure \ref{fig:age_aids_vl} contains the estimated conditional quantiles and probabilities as functions of whether a patient had an AIDS event prior to starting ART. People with a prior AIDS event (36\%) tended to have a higher VL at 6 months.  The third row of Figure \ref{fig:age_aids_vl} is the estimated conditional quantiles and probabilities as functions of baseline VL.  People with a higher baseline VL tended to have a higher VL at 6 months.  

Supplementary Material S2.2 contains results from a similar CPM, except with continuous covariates expanded using restricted cubic splines to relax linearity assumptions and increase model flexibility. The results are fairly similar.

For comparisons, we also analyzed the data using competing approaches described earlier. First, we fit logistic regression to 6-month VL values dichotomized as ${<}400$ vs. ${\ge}400$ copies/mL, corresponding to the highest DL. Results are in Table S4 of the Supplementary Material S2.2. The CPM and the logistic regression model gave similar estimates of the beta coefficients (which are log odds ratios), although there were some differences in the estimates and the CIs from CPMs tend to be narrower, as expected. 
In logistic regression, the log odds ratios are based on the single undetectable vs. detectable dichotomization, while those in CPMs are based on dichotomizations at each response value. Second, we fit a full likelihood-based model 
assuming the outcome variable was normally distributed after $\log_{10}(\cdot)$ transformation.  Note that even the log-transformed 6-month VL was still quite skewed (shown in Figure S4), and hence the assumptions of this fully parametric approach were questionable. The parameters in this approach and those from the CPM are not directly comparable because they are on different scales, however, the directions of associations were similar. 

\section{Simulations}

Extensive simulations of CPMs with continuous data have been reported elsewhere \citep{liu2017modeling, tian2020empirical}. Here we present a limited set of simulations investigating the performance of CPMs with data subject to single and multiple DLs.

\subsection{Single Detection Limits}

Data were generated for sample sizes of $n=100$ and $n=500$ such that the outcome $Y$ followed a normal linear model after log transformation in the following manner:
\begin{align*}
    Y = \exp(Y^*), \textrm{ where } Y^* &= X\beta + \epsilon, 
    \beta =1, 
    X \sim N(0,1), \textrm{ and }
    \epsilon \sim N(0,1).
\end{align*}
Various scenarios of DLs of $Y$ were considered: 1. No DL.  2. One lower DL at 0.25 (censoring rate 16.3\%). 3. One upper DL at 4 (censoring rate 16.3\%). 4. One lower DL at 0.25 and one upper DL at 4 (censoring rate 32.7\%). 5. One lower DL at 4 (censoring rate 83.7\%).
In addition, we considered a scenario with a more complicated transformation:
6. One lower DL at 0.0625 and
\begin{equation*}
    Y = \left\{\begin{array}{lr}
    \exp(2Y^*) \textrm{ if } Y^*<\log(0.25)\\ 
    \sqrt{\exp(Y^*)}\textrm{ if } \log(0.25) \le Y^* <\log(2)\\ 
    \exp(Y^*)\textrm{ if } Y^*\ge \log(2).
    \end{array} \right.
\end{equation*}
Note that the $Y$ in scenario 6 is a monotonic transformation of that in scenario 2 with exactly the same censoring rate.

CPMs were fit to the observed data $\{X,Y\}$ without any knowledge of the correct transformation or $Y^*$.  We simulated 10,000 replications under each scenario. Percent bias, root mean squared error (RMSE), and coverage of 95\% CIs were estimated with respect to $\beta$, conditional medians for $X=\{0,1\}$, and conditional CDFs at $y=1.5$ for $X=\{0,1\}$.

Table \ref{tab:one_dl} shows results under correctly specified models (i.e., probit link function and $X$ correctly included).  CPMs resulted in nearly unbiased estimation and good CI coverage.  As the sample size increased, both the bias and RMSE decreased. 
Note that estimation of the condition medians was ``perfect'' in scenario 5 because the true conditional medians were below the lower DL due to the high censoring rate and the estimated conditional medians were always `${<}4$', the lowest outcome category corresponding to below the DL.  The estimate of $\beta$ was more variable in scenario 5 because of the high censoring rate. 
The estimation of $\beta$ in scenario 6, where data were generated from the complicated transformation, was exactly the same as that in scenario 2 because the same seed was used in all simulation scenarios and the order information above the DL was identical between scenarios 2 and 6. However, the conditional medians and CDFs depend on the scale of the outcome, and their estimates differed between scenarios 2 and 6.

\begin{table}
\begin{threeparttable}
    \centering
    \caption{Simulation results for single DLs}
    \label{tab:one_dl}
  
    \renewcommand{\arraystretch}{0.53} 
    \begin{tabular}{@{\extracolsep{5pt}}cccccccc@{}}
    \hline
     & & \multicolumn{3}{c}{n=100} & \multicolumn{3}{c}{n=500}\\
     \cline{3-5}\cline{6-8}
  Parameter & Truth & Bias(\%) & RMSE & Coverage  &   Bias(\%)
  & RMSE & Coverage\\
  \hline
    \textbf{Scenario 1} & & & & & & & \\
    $\beta$ & 1 & 2.803 & 0.133 & 0.944 & 0.638 & 0.057 & 0.945\\
    $Q(0.5|X=0)$ &  1 & -0.388 & 0.140 & 0.951 & -0.124 & 0.063 & 0.951\\
    $Q(0.5|X=1)$ & 2.718 &  1.552 & 0.494 & 0.949 & 0.321 & 0.218 & 0.951\\ 
    $F(1.5|X=0)$ &  0.658 &  0.117 & 0.054 & 0.949 & 0.059 & 0.024 & 0.951\\
    $F(1.5|X=1)$ & 0.276 & -1.429 & 0.060 & 0.949 & -0.383 & 0.026 & 0.945\\ 
     \hline
    \textbf{Scenario 2}& & & & & & & \\
    $\beta$ & 1& 2.665 & 0.138 & 0.945 & 0.585 & 0.057 & 0.948\\
    $Q(0.5|X=0)$ & 1 &  -0.240 & 0.142 & 0.953 & 0.028 & 0.063 & 0.948\\
    $Q(0.5|X=1)$ & 2.718 & 1.445 & 0.498 & 0.953 & 0.406 & 0.222 & 0.946\\
    $F(1.5|X=0)$ & 0.658 & 0.005 & 0.054 & 0.946 & -0.085 & 0.024 & 0.950\\
    $F(1.5|X=1)$ & 0.276 & -0.479 & 0.061 & 0.948 & 0.368 & 0.028 & 0.943\\
     \hline
    \textbf{Scenario 3}& & & & & & &\\
    $\beta$ & 1 & 2.710 & 0.139 & 0.943 & 0.581 & 0.058 & 0.948\\
    $Q(0.5|X=0)$ & 1 & -0.460 & 0.141 & 0.951 & -0.020 & 0.063 & 0.949\\
    $Q(0.5|X=1)$ & 2.718 & 0.803 & 0.477 & 0.954 & 0.310 & 0.223 & 0.945\\
    $F(1.5|X=0)$ & 0.658 & 0.0147 & 0.054 & 0.946 & -0.083 & 0.024 & 0.951\\
    $F(1.5|X=1)$ & 0.276 & -0.487 & 0.062 & 0.948 & 0.381 & 0.028 & 0.941\\
     \hline
    \textbf{Scenario 4}& & & & & & &\\
    $\beta$ & 1 & 2.544 & 0.139 & 0.945 & 0.538 & 0.058 & 0.951\\
    $Q(0.5|X=0)$ & 1 & -0.243 & 0.141 & 0.954 & 0.028 & 0.063 & 0.949\\
    $Q(0.5|X=1)$ & 2.718 & 1.017 & 0.477 & 0.953 & 0.358 & 0.223 & 0.947\\
    $F(1.5|X=0)$ & 0.658 & 0.004 & 0.054 & 0.947 & -0.086 & 0.024 & 0.950 \\
    $F(1.5|X=1)$ & 0.276 & -0.285 & 0.062 & 0.948 & 0.432 & 0.028 & 0.943\\
     \hline
    \textbf{Scenario 5}& & & & & & \\
    $\beta$ & 1 & 7.315 & 0.276 & 0.946 & 1.330 & 0.101 & 0.948\\
    $Q(0.5|X=0)$ & 1 & 0\tnote{*} & 0 & 1 & 0 & 0 & 1\\
    $Q(0.5|X=1)$ & 2.718 & 0 & 0 & 1 & 0 & 0 & 1\\
    $F(1.5|X=0)$ & 0.658 & 0.183 & 0.026 & 0.954 & -0.029 & 0.010 & 0.949 \\
    $F(1.5|X=0)$ & 0.276 &-0.189 & 0.069 & 0.952 & -0.169 & 0.030 & 0.949 \\  
  \hline
  \textbf{Scenario 6} & & & & & & \\
    $\beta$ & 1 & 2.665 & 0.138 & 0.945 & 0.585 & 0.057 & 0.948 \\
    $Q(0.5|X=0)$ & 1 & -0.841 & 0.071 & 0.951 & -0.503 & 0.032 & 0.945\\
    $Q(0.5|X=1)$ & 0.368 & -0.312 & 0.542 & 0.953 & -0.529 & 0.222 & 0.946\\
    $F(1.5|X=0)$ & 0.654 & 0.254 & 0.048 & 0.947 & 0.056 & 0.022 & 0.953\\
    $F(1.5|X=1)$ & 0.500 & 0.061 & 0.069 & 0.949 & 0.536 & 0.032 & 0.946\\
    \hline  
          
    \end{tabular}
    
\begin{tablenotes}\footnotesize
\item[*] The results of zero bias and RMSE when there is a high censoring rate are because the true conditional medians are below the lower DL and the estimated conditional medians were always `${<}4$', the lowest outcome category corresponding to below the DL.
\end{tablenotes}

\end{threeparttable}
\end{table}

Table S5 in the Supplementary Material shows results under scenario 2 with $n=1000$ comparing CPMs with some widely used methods for handling DLs, specifically single imputation with $l/2$, single imputation with $l/\sqrt{2}$, multiple imputation, and fully parametric maximum likelihood estimation (MLE).  
For all non-CPM approaches, we first correctly assumed that the outcome variable followed a log-normal distribution.  With the imputation approaches, unobserved values were imputed, then a linear regression model was fit on the log-transformed outcome to obtain the $\beta$ estimate, and median regression was used to estimate conditional medians.  In multiple imputation, the correct tail distribution was used for imputing data and 10 iterations were performed for each data set. 
As expected, the MLE performed the best with the lowest bias and RMSE, and highest efficiency because the distributional assumptions matched the true distribution. 
The performance of multiple imputation was similar to that of the MLE, but with higher RMSE. As a semiparametric method, the CPM,
also resulted in minimal bias and correct coverage but had slightly larger variance and RMSE. In contrast, the single imputation estimators were biased and tended to have poor coverage, especially for estimating $\beta$.
We also evaluated the comparator methods under misspecification of the transformation. We simulated datasets with $X\sim N(5,1)$, $Y^*=X\beta + \epsilon$, $\beta=1$, $\epsilon \sim N(0,1)$, $Y = {Y^*}^2$, $n=1000$, and $l=13.12$ (approximately 17\% censored). The non-CPM approaches assumed a normal linear model after an incorrectly specified log-transformation.  As shown in the bottom half of Table S5, only the CPM was able to properly estimate $\beta$ and the conditional medians, because  pre-transformation and strict distributional assumptions are not needed for fitting CPMs.

Finally, the Supplementary Material Table S6 shows the performance of CPMs for the data generated in scenario 2 under moderate and severe link function misspecification (i.e., fitting CPMs with logit and loglog link functions, respectively).
Link function misspecification is equivalent to misspecification of the distribution of $\epsilon$ because $F_\epsilon=G^{-1}$. 
The performance of CPMs was reasonable with moderate link function misspecification with bias under 6\% and coverage of 95\% CI close to 0.95 with $n=100$, although as low as 0.91 with $n=500$. With severe link function misspecification, the performance of CPMs was noticeably worse, with bias as high as 12\% and coverage as low as 0.60 for the conditional median at $X=1$.

\subsection{Multiple Detection Limits}

To illustrate the use of CPMs with multiple detection limits, we simulated data from 3 study sites.  The data were generated in a similar way as in Section 4.1, but different DLs were applied at different sites and the distribution of the covariate $X$ was allowed to vary across sites in some scenarios.  Specifically, we considered the following 5 scenarios:
\begin{enumerate}
    \item[1.] Lower DLs 0.16, 0.30, and 0.50 for the 3 sites (about 10\%, 20\%, and 30\% censored), and $X$ is independent of DLs/sites.
    \item[2.] Upper DLs 0.16, 0.30, and 0.50 for the 3 sites (about 90\%, 80\%, and 70\% censored), and $X$ is independent of DLs/sites.
    \item[3.] Lower DLs 0.16, 0.30, and 0.50 for the 3 sites (about 17\%, 20\%, and 20\% censored), and $X \sim N(\mu_x,1)$ where $\mu_x=-0.5, 0,$ and 0.5 for site 1, 2, and 3, respectively.
    \item[4.] Upper DLs 0.16, 0.30, and 0.50 for the 3 sites (about 83\%, 80\%, and 80\% censored), and $X \sim N(\mu_x,1)$ where $\mu_x=-0.5, 0,$ and 0.5 for site 1, 2, and 3, respectively.
    \item[5.] Lower DLs 0.2, 0.3, and -$\infty$ (13\%, 20\%, and 0\% censored) and upper DLs at $\infty$, 4, and 3.5 (0\%, 19\%, and 16\% censored) for the 3 sites, and $X$ is independent of DLs/sites.
\end{enumerate}
We considered two sample sizes, $n=150$ and $n=900$, with the sample sizes distributed equally across sites. In scenarios 2 and 4, because of the high censoring rates, we estimated the quantiles at $p=0.03$ (i.e., 3rd percentile) and CDFs at $y=0.05$.  Results from fitting the CPM based on 10,000 replications are shown in Supplementary Material S3. In summary, estimates had very low bias and confidence intervals had proper coverage in all simulation scenarios.

\section{Discussion}

In this paper, we have described an approach to address detection limits in response variables using CPMs. CPMs have several advantages over existing methods for addressing DLs. They make minimal distributional assumptions, they yield interpretable parameters, and they are invariant to the value assigned to measures outside DLs. Any values outside the lowest/highest DLs are simply assigned to the lowest/highest ordinal categories, and estimation proceeds naturally. CPMs are also easily extended to handle multiple DLs.  From simulation studies, we saw that CPMs performed well, even with high censoring rates and relatively small sample sizes. We also illustrated the use of CPMs with two quite different HIV datasets with censored response data. Similar datasets with limits of detection are quite common in biomedical research; the CPM is an effective analysis tool in these settings.

CPMs have some limitations. Although CPMs do not make distributional assumptions on the response variable, the link function must still be specified, which corresponds to making an assumption on the distribution of the response variable after an unspecified transformation. Performance can be poor with severe link function misspecification; however, CPMs appear to be fairly robust to moderate misspecification. In addition, because we do not make distributional assumptions outside DLs, we are not able to estimate conditional expectations after fitting a CPM; however, with DLs, conditional quantiles are probably more reasonable statistics to report anyway. The codes for applications and simulations are available at https://github.com/YuqiTian35/DetectionLimitCode.

Further research could consider extensions of CPMs to handle clustered or longitudinal data with DLs. It may be of interest to study the use of these models with right-censored failure time data (i.e., survival data), where each observation is  potentially subject to a different censoring time; the current manuscript only considered situations with a relatively small number of potential censoring times (i.e., upper DLs).

\bibliographystyle{apalike}
\bibliography{main}

\end{document}


\def\spacingset#1{\renewcommand{\baselinestretch}%
{#1}\small\normalsize} \spacingset{1}

\if1\blind
{
  \title{\bf  Supplementary Material to ``Addressing Detection Limits with Semiparametric Cumulative Probability Models''}
  \author{Yuqi Tian\\
  
    Department of Biostatistics, Vanderbilt University,\\
    Chun Li \\
    Department of Population and Public Health Sciences, \\University of Southern California,\\
    Shengxin Tu \\
    Department of Biostatistics, Vanderbilt University,\\
    Nathan T. James \\
    Department of Biostatistics, Vanderbilt University,\\
    Frank E. Harrell \\
    Department of Biostatistics, Vanderbilt University,\\
    and \\
    Bryan E. Shepherd\\
    Department of Biostatistics, Vanderbilt University}
  \maketitle
} \fi

\if0\blind
{
  \bigskip
  \bigskip
  \bigskip
  \begin{center}
    {\LARGE\bf Supplementary Material to ``Addressing Detection Limits with Semiparametric Cumulative Probability Models''}
\end{center}
  \medskip
} \fi

\bigskip

\renewcommand{\thesection}{S\arabic{section}}
\renewcommand{\thefigure}{S\arabic{figure}}  
\renewcommand{\thetable}{S\arabic{table}}  

\spacingset{1.9} 

\section{Methods}
\subsection{Cumulative Probability Models}

Let $a_1<\cdots<a_J$ be the unique outcome values in a data set.  In a CPM,
\begin{equation*}
  G[\Pr(Y\le a_j|X, \alpha, \beta)] =\alpha_j-\beta^T X \quad (j=1,...,J-1).
\end{equation*}
For observation $i$ with $X_i=x_i$, let
$\gamma_{i,j} =\Pr(Y_i\le a_j|x_i, \alpha, \beta)$.  For convenience, let
$\gamma_{i,0}=0$ and $\gamma_{i,J}=1$.  Let
$p_{i,j}=\gamma_{i,j}-\gamma_{i,j-1}$ be the multinomial probability at $a_j$
($j=1,\ldots,J$).

Consider the logit link function, $G(p)=\log \left(p / (1-p)\right)$.  Then for
$j=1,\ldots,J-1$, $\gamma_{i,j}=\exp(\phi_{i,j}) \left(1+\exp(\phi_{i,j})\right)$,
where $\phi_{i,j}=\alpha_j-\beta^T x_i$, and
$\partial\gamma_{i,j} / \partial\beta
=\left(\partial\gamma_{i,j} / \partial\phi_{i,j}\right)
\left(\partial\phi_{i,j} / \partial\beta\right) =-x_i\gamma_{i,j}(1-\gamma_{i,j})$.
Note that this equation also holds for $j=0$ and $j=J$.  Then
\begin{align*}
  \frac{\partial p_{i,j}}{\partial\beta}
  =\frac{\partial\gamma_{i,j}}{\partial\beta}
  -\frac{\partial\gamma_{i,j-1}}{\partial\beta}
  &=-x_i[\gamma_{i,j}(1-\gamma_{i,j}) -\gamma_{i,j-1}(1-\gamma_{i,j-1})] \\
  &=-x_i[(\gamma_{i,j-1}+p_{i,j})(1-\gamma_{i,j}) -\gamma_{i,j-1}(1-\gamma_{i,j}+p_{i,j})] \\
  &=-x_i[p_{i,j}(1-\gamma_{i,j}) -\gamma_{i,j-1}p_{i,j}] \\
  &=x_ip_{i,j}(\gamma_{i,j}+\gamma_{i,j-1}-1).
\end{align*}

Given data $\{(x_i,y_i)\}$, the log-likelihood is
$l(\alpha,\beta) =\sum_i \log p_{i,j(i)}$, where $j(i)$ is the index such
that $a_{j(i)}=y_i$.  The score function with respect to $\beta$ is
\begin{equation*}
  \frac{\partial l}{\partial\beta} =\sum_i\frac{1}{p_{i,j(i)}} \frac{\partial
    p_{i,j(i)}}{\partial\beta} =\sum_i x_i(\gamma_{i,j(i)}+\gamma_{i,j(i)-1}-1).
\end{equation*}

Now consider the situation where there is a single binary $X$, coded as $0$
and $1$.  Let $n_k$ be the number of observations with $X=k$ ($k=0,1$), and
$n=n_0+n_1$.  Under the null of $\beta=0$, the CPM estimate of $\alpha_j$ is
$\hat\alpha_j = G(\hat P_j)$, where $\hat P_j$ is the CDF of the empirical
distribution of $\{y_i\}$ at $a_j$.  As a result,
$\hat\gamma_{i,j(i)}=\hat P_{j(i)}$.  In this situation, the numerator of the
score test statistic is
\begin{equation*}
  S= \frac{\partial l}{\partial\beta}\bigg\rvert_{\hat\alpha|\beta=0}
  =\sum_{i:x_i=1} (\hat P_{j(i)}+\hat P_{j(i)-1}-1)
  =\sum_{i:x_i=1} (\hat P_{j(i)}+\hat P_{j(i)-1})-n_1.
\end{equation*}

Let $k_j$ be the number of observations with $y_i=a_j$, $R_j$ be the midrank
for $a_j$, and $R_1 =\sum_{i:x_i=1} R_{j(i)}$ be the sum of the midranks for
the observations with $X=1$.  Then
$R_j =n\hat P_j- (k_j-1) / 2 =n\hat P_{j-1}+ (k_j+1) / 2$.  Since $ n (\hat P_{j(i)}+\hat P_{j(i)-1}) / 2 =
(R_{j(i)}+ (k_j-1)/2 +R_{j(i)} -(k_j+1)/2) /2 =R_{j(i)}- 1/2$,
we have
\begin{equation*}
  \frac{n}{2}S =R_1-\frac{n_1}{2}-\frac{nn_1}{2} =R_1-\frac{n_1(n+1)}{2}.
\end{equation*}
Note that $E(R_1)=n_1(n+1) / 2$ under the null of $\beta=0$.

In comparison, the Wilcoxon--Mann--Whitney statistic is
$U=R_1-n_1(n_1+1) / 2$, which has mean $\mu_U=n_1n_0 / 2$.  The
corresponding test statistic, $z= (U-\mu_U)/\sigma_U$, has numerator
$U-\mu_U =R_1-n_1(n+1)/2$.  Thus the numerator of the Wilcoxon test
statistic differs from that of the score test statistic by a constant
multiplier $n/2$.

\subsection{Multiple Detection Limits}

We show a toy example in Table \ref{tab:example} to illustrate the approach to handle multiple DLs described in Section 2.3.

\begin{table}
     \caption{The likelihood contribution for each observation in a toy example with multiple detection limits. Suppose there are eight observations from two sites. The lower and upper DLs of one site are 3 and 9, respectively, and those of the other site are 5 and 12. Suppose the dataset is 
     $\{(3, L; x_1), (4, 1; x_2), (6, 1; x_3), (9,U;x_4)\}$ from site 1 and $\{(5,L;x_5), (7,1;x_6), (10,1;x_7), (12,U;x_8)\}$ from site 2. There are thus $J=4$ uncensored values, and we include two additional categories corresponding to those below the smallest DL and those above the largest DL, such that 
     $S=\{a_0, a_1, \dots, a_5\}=\{\text{`}{<}3\text{'}, 4, 6, 7, 10, \text{`}{>}12\text{'}\}$.}
     \label{tab:example}
    \begin{center}
    \renewcommand{\arraystretch}{0.53} 
    \begin{tabular}{cccc}
    \hline
        $z_i$ & $\delta_i$ & $j$ & Likelihood  \\
    \hline
    3 & $L$ & 0 & $F_\epsilon(\alpha_0-\beta^T x_1)$ \\
    4 & 1 & 1 & $F_\epsilon(\alpha_1-\beta^T x_2)-F_\epsilon(\alpha_0-\beta^T x_2)$ \\
    5 & $L$ & 1 & $F_\epsilon(\alpha_1-\beta^T x_5)$ \\
    6 & $1$ & 2 & $F_\epsilon(\alpha_2-\beta^T x_3)-F_\epsilon(\alpha_1-\beta^T x_3)$ \\
    7 & 1 & 3 & $F_\epsilon(\alpha_3-\beta^T x_6)-F_\epsilon(\alpha_2-\beta^T x_6)$ \\
    9 & $U$ & 4 & $1-F_\epsilon(\alpha_3-\beta^T x_4)$ \\
    10 & 1 & 4 & $F_\epsilon(\alpha_4-\beta^T x_7)-F_\epsilon(\alpha_3-\beta^T x_7)$ \\
    12 & $U$ & 5 & $1-F_\epsilon(\alpha_4-\beta^T x_8)$\\
 \hline
    \end{tabular}
\end{center}
\end{table}

\section{Applications}

\subsection{Single Detection Limit}

\begin{figure}[!htbp]
    \centering
    \includegraphics[scale=0.7]{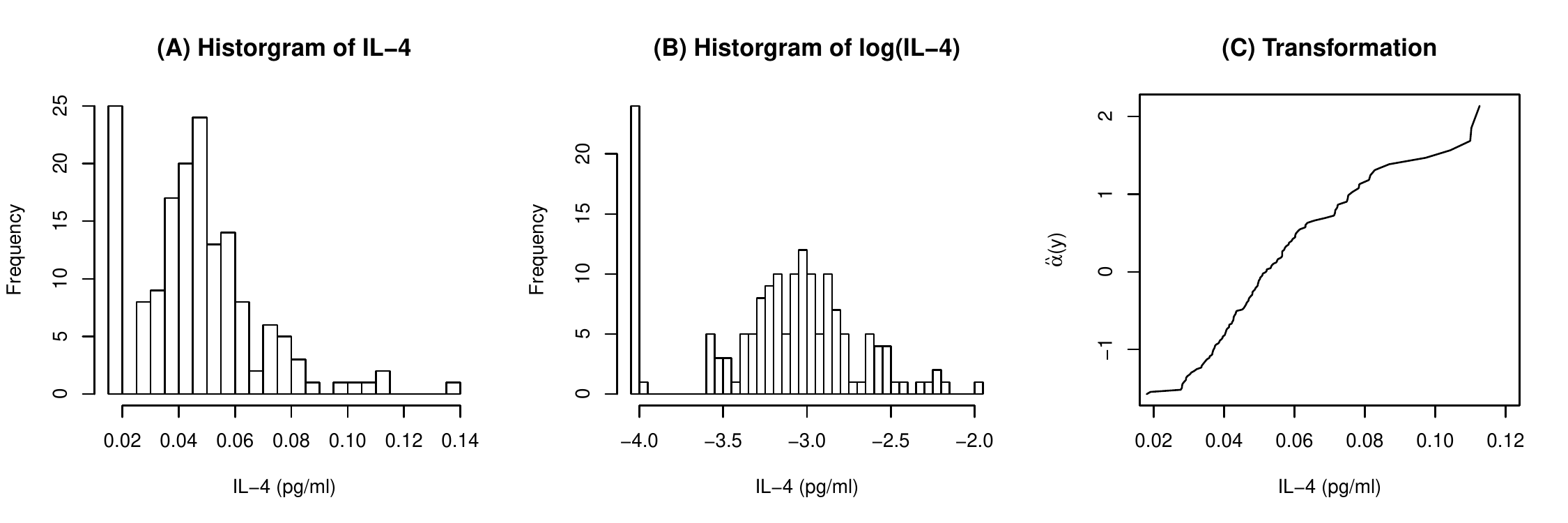}
    \caption{(A) The distribution of IL-4. (B) The distribution of log-transformed IL-4. (C) The estimated transformation function.}
    \label{fig:il4}
\end{figure}

\begin{figure}
    \centering
    \includegraphics[scale=0.75]{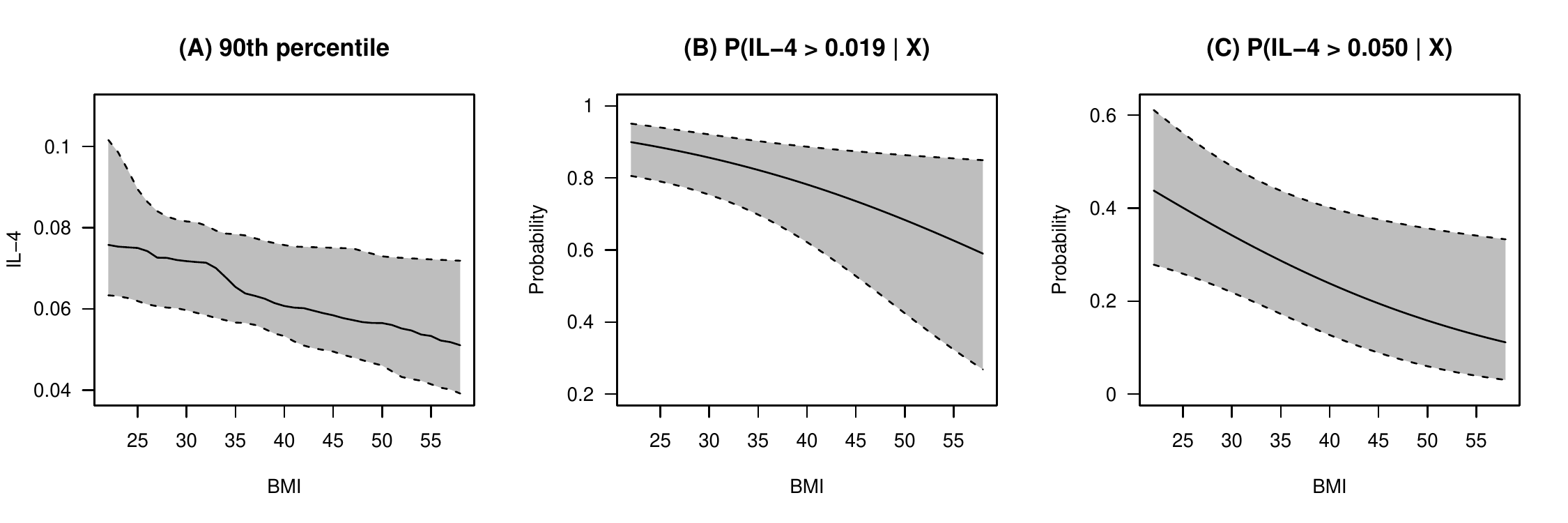}
    \caption{We can obtain conditional quantities by CPMs. Other covariates set to corresponding median or mode.  (A) The conditional 90th percentile of IL-4 as a function of BMI. (B)  The probabilities of IL-4 being greater than 0.019 (the DL). (C) The probabilities of IL-4 being greater than 0.05.}
    \label{fig:single_conditional_quantities}
\end{figure}

Figure \ref{fig:il4} shows that the distribution of IL-4 is right-skewed and has a lower detection limit of 0.019. After log-transformation, the response variable is approximately normally distributed except for values below the DL. The results for fitting the CPM is shown in Table \ref{tab:cpm_single} and the estimated transformation function is in Figure \ref{fig:il4}. Based on the results of CPMs, we can derive conditional quantities. For example, Figure \ref{fig:single_conditional_quantities} shows the 90th percentile of IL-4 as a function of BMI, and the probabilities of IL-4 being greater than 0.019 (the DL) and greater than 0.05 as functions of BMI while setting other covariates to the median (for continuous covariates) or mode (for categorical covariates).

\begin{table}[]
    \centering
    \caption{We show odds ratio with 95\% confidence intervals, and p-values of parameter estimation for predictors included in the model.}
    \label{tab:cpm_single}
    \renewcommand{\arraystretch}{0.53} 
    \begin{tabular}{lcccc}
        \hline
        Predictor & Odds Ratio & 95\% (CI) & P-value & PI\\
        \hline
        \textbf{Age} (per 10 years) &	1.16 &  (0.89, 1.51)	& 0.271 & 0.525\\
        
        \textbf{Sex} & & & 0.689 & \\
        \hspace{3mm} Female (reference) & 1 & & \\
        \hspace{3mm} Male & 1.14 & (0.59, 2.22) & 	0.689 & 0.522\\
        
        \textbf{BMI} (per 5 kg/$\text{m}^2$) &	0.78  & (0.62, 0.97) & 0.023 & 0.264\\
        
        \textbf{Status} & & & ${<}0.001$ & \\
        \hspace{3mm} HIV positive, insulin sensitive (reference) & 1 & & \\
        \hspace{3mm}  HIV positive, pre-diabetic   & 2.15 &  (1.03, 4.50) & 0.041 & 0.625\\
        \hspace{3mm} HIV positive, diabetic  &	1.03 & (0.44,	2.44) & 	0.945 & 0.505\\
        \hspace{3mm} HIV negative, diabetic  & 1.82 &(0.70,	4.74) &	0.218 & 0.599\\
    
        \hline
    \end{tabular}
\end{table}

\begin{figure}
    \centering
    \includegraphics[scale=0.75]{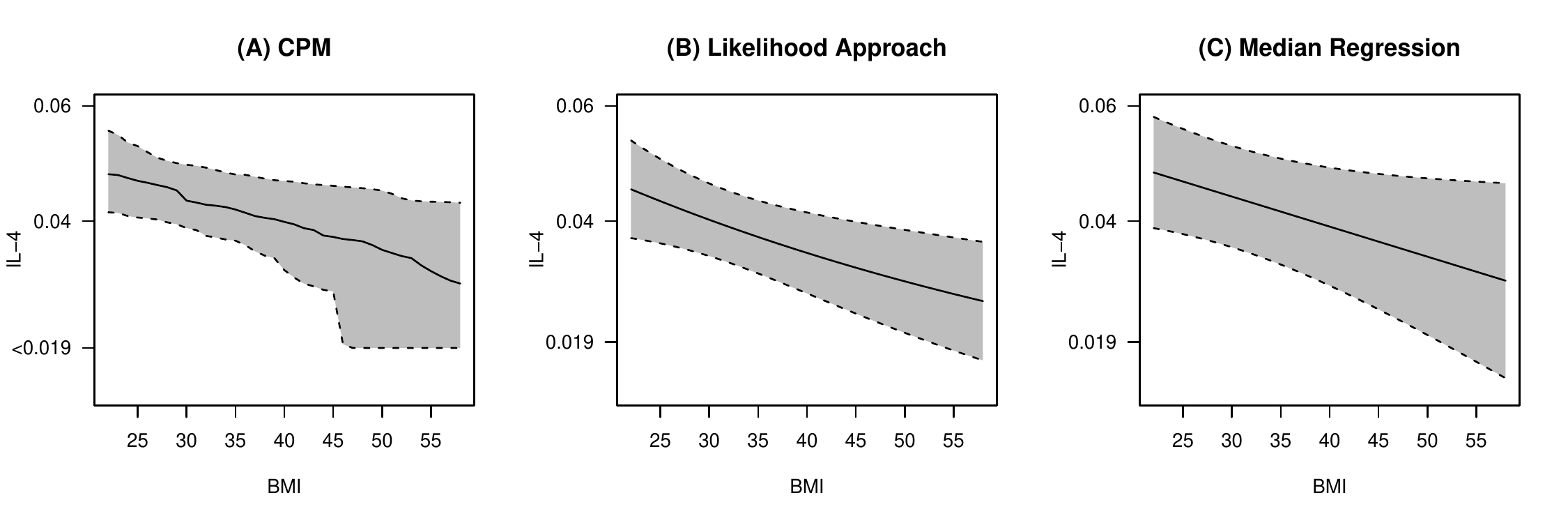}
    \caption{(A) The conditional median obtained by CPMs fixing other covariates. (B)  The conditional median obtained by the likelihood approach (after log-transformation). (C)  The conditional median obtained by median regression.}
    \label{fig:median_logit}
\end{figure}

We fit median regression and the likelihood approach to the data and obtain the conditional medians. The comparison of the conditional medians is in Figure \ref{fig:median_logit}.

\subsection{Multiple Detection Limits}

\begin{figure}
    \centering
    \includegraphics[scale=0.6]{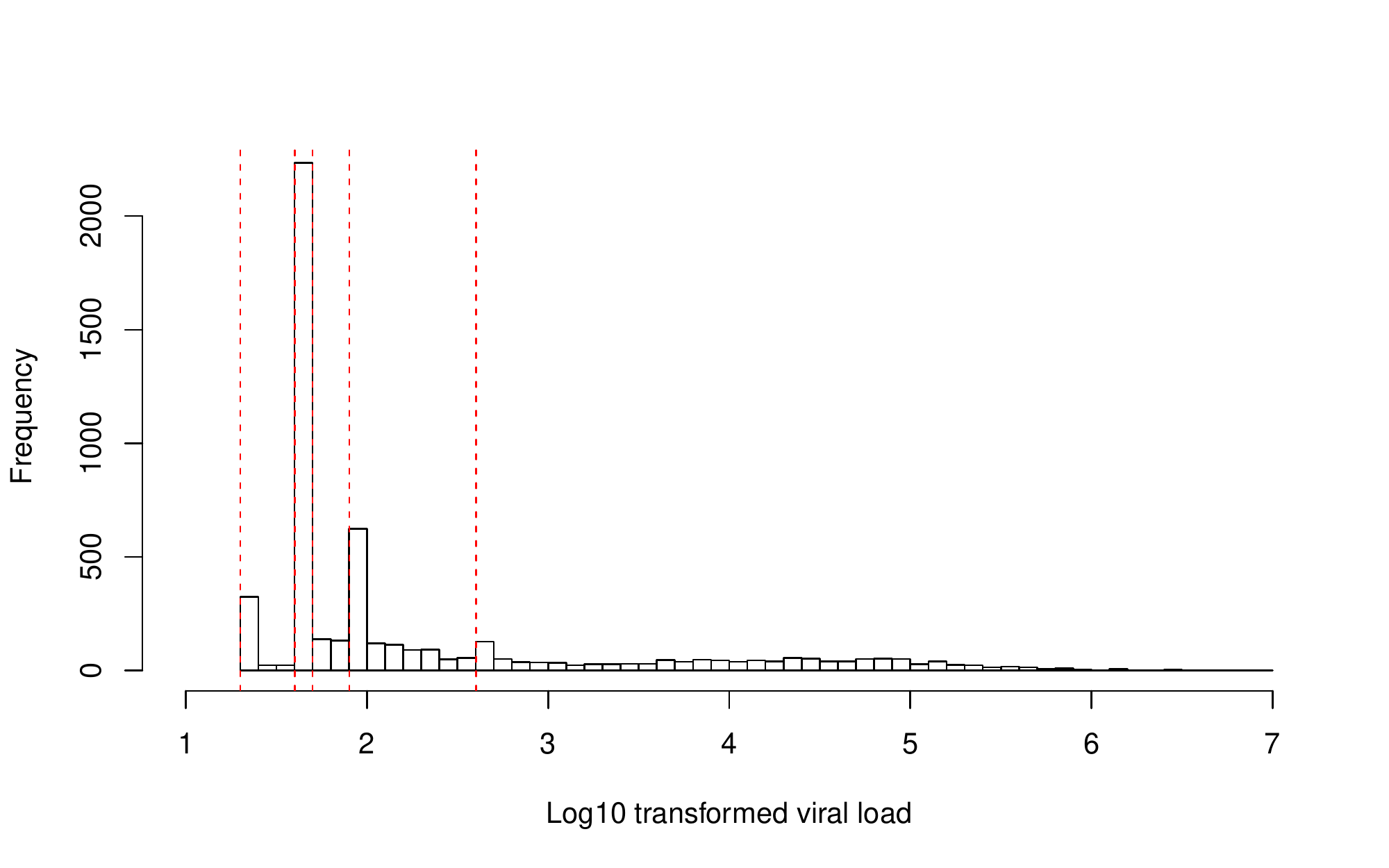}
    \caption{Distribution of the log10 transformed 6-month VL. The transformed DLs are shown in red dashed lines; 2992 (56\%) of the patients were censored at one of these DLs.}
    \label{fig:log10trans}
\end{figure}

Figure \ref{fig:log10trans} shows the distribution of  log-transformed 6-month VL and lower DLs.

\label{s:appendix_a}
\begin{figure}
    \centering
    \includegraphics[scale=0.7]{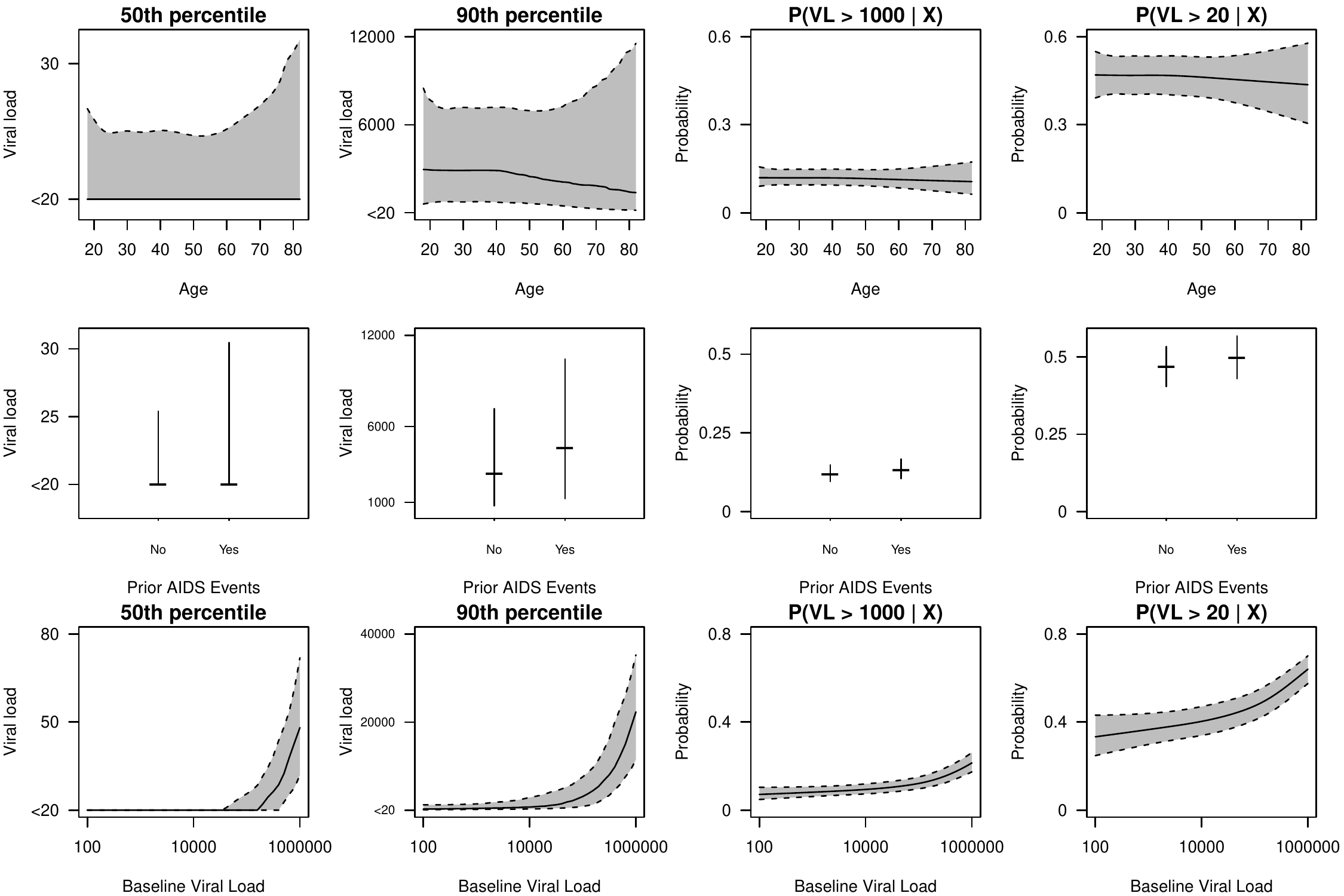}
    \caption{The estimated conditional 50th and 90th percentile of age, prior AIDS events, and 6-month viral load and the conditional probability of the 6-month viral load being greater than 1000 and 20 as functions of age, prior AIDS event, and baseline viral load while keeping other covariates at their medians (for continuous variables) or modes (for categorical variables) level based on the model with splines.}
    \label{fig:splines_all}
\end{figure}

\begin{center}
\renewcommand{\arraystretch}{0.53} 
\begin{longtable}{lccc}
\caption{The association between the outcome and predictors by odds ratios and their 95\% confidence. P-values are based on likelihood ratio tests. Baseline CD4 values are square root transformed. Baseline viral load are log (base 10) transformed. Time indicates time from ART initiation to 6-month viral load measurement.} \label{tab:res_splines} \\

\hline Predictor & Odds Ratio & 95\% CI & P-value \\  \hline 
\endfirsthead

\multicolumn{4}{c}%
{{\bfseries \tablename\ \thetable{} -- continued from previous page}} \\
\hline Predictor & Odds Ratio & 95\% CI & P-value \\ \hline 
\endhead

\hline \multicolumn{4}{r}{{Continued on next page}} \\ \hline
\endfoot

\hline \hline
\endlastfoot

\textbf{Age} & & & 0.959 \\
\hspace{3mm}20 & 1.00 & 0.82-1.23 & \\
\hspace{3mm}30  (reference) & 1 &  & \\
\hspace{3mm}40 & 1.00 &  0.87-1.14 &\\
\hspace{3mm}50 & 0.98 &  0.84-1.13 &\\
\hspace{3mm}60 & 0.94 &  0.76-1.16 &\\

\textbf{Sex} & & & 0.213 \\
\hspace{3mm} Male  (reference) & 1 &  & \\
\hspace{3mm} Female & 0.90 & 0.76-1.06 &\\

\textbf{Study center} &   & & ${<}0.001$ \\
\hspace{3mm} Peru  (reference)  & 1 &  & \\
\hspace{3mm} Argentina  & 1.17 &  0.91-1.50 &\\
\hspace{3mm} Brazil  & 0.99 &  0.83-1.17 &\\
\hspace{3mm} Chile   & 0.96 &  0.81-1.15 &\\
\hspace{3mm} Mexico  & 0.57 &  0.47-0.68&\\

\textbf{Route of infection} & & & 0.099 \\
\hspace{3mm} Homosexual/Bisexual  (reference)  & 1 &  & \\
\hspace{3mm} Heterosexual &  0.93 &  0.81-1.07 & \\
\hspace{3mm} Other/Unknown & 0.77 & 0.60-0.98 & \\

\textbf{Prior AIDS event} & & & 0.075\\
\hspace{3mm} No  (reference) & 1 &  & \\
\hspace{3mm} Yes & 1.13 &  0.99-1.29 & \\

\textbf{Baseline CD4} & & & ${<}0.001$ \\
\hspace{3mm}50 & 0.66 &  0.60-0.73 & \\
\hspace{3mm}100 & 0.65 &  0.60-0.69 & \\
\hspace{3mm}200  (reference) &  1 &  & \\
\hspace{3mm}300 & 1.71 & 1.59-1.85 & \\

\textbf{Baseline VL} & & & ${<}0.001$ \\
\hspace{3mm}100 & 0.74 &  0.52-1.04 & \\
\hspace{3mm}1000 & 0.85 &  0.73-1.01 & \\
\hspace{3mm}10000  (reference) &  1 &  & \\
\hspace{3mm}100000 & 1.33 &  1.18-1.50 &\\

\textbf{ART regimen} & & & ${<}0.001$ \\
\hspace{3mm} NNRTI-based  (reference) &  1 &  & \\
\hspace{3mm} INSTI-based & 0.60 &  0.43-0.83   &\\
\hspace{3mm} PI-based & 111 &  0.95-1.29 &\\
\hspace{3mm} Other & 2.51 &  1.26-5.00 &\\

\textbf{Time} & & & ${<}0.001$ \\
\hspace{3mm}3 months & 1.57 &  1.20-2.04 &\\
\hspace{3mm}6 months  (reference) &  1 &  & \\
\hspace{3mm}9 months & 1.05 &  0.91-1.22 & \\

 \textbf{Calendar year} & & & ${<}0.001$ \\
\hspace{3mm}2005 & 1.50 &  1.33-1.68 &\\
\hspace{3mm}2010  (reference) &  1 &  & \\
\hspace{3mm}2015 & 0.45 &  0.38-0.52 & \\
\end{longtable}
\end{center}

\begin{table}[]
    \centering
    \caption{The odds ratio analyzing the data by logistic regression and point estimation obtained by the fully likelihood approach. For logistic regression,  the responses are dichotomized into two levels: ${<}400$ and ${\ge}400$. We show odds ratios with 95\% confidence intervals, and p-values of parameter estimation for predictors included in the model. The point estimation with 95\% confidence intervals, and p-values of parameter estimation for predictors included in the model by the fully likelihood approach.}
    \label{tab:logistic}
    \renewcommand{\arraystretch}{0.53} 
    \begin{tabular}{@{\extracolsep{5pt}}lcccc@{}}
         \hline
    & \multicolumn{2}{c}{Logistic Regression} & \multicolumn{2}{c}{Fully Likelihood Approach}\\
     \cline{2-3}\cline{4-5}
  & Odds Ratio (CI) & P-value & Estimation (CI) & P-value\\
  \hline
  
        \textbf{Age} (per 10 years) &	0.91 (0.85, 0.98)	& 0.013 &	-0.03 (-0.09, 0.03)	& 0.276\\
        
        \textbf{Sex} & & 0.739 & & 0.271 \\
        \hspace{3mm} Male (reference) & 1 & & 1 & \\
        \hspace{3mm} Female & 0.96	(0.77, 1.21) & -0.10 &	(-0.28, 0.08) &	\\
        
        \textbf{Study center} & & ${<}0.001$ & & ${<}0.001$\\
        \hspace{3mm} Peru (reference) & 1 &   & 1 &  \\
        \hspace{3mm} Argentina  &	1.43 (1.04,	1.97) &  & 0.26  (-0.01, 0.53) & \\
        \hspace{3mm} Brazil  &	1.21 (0.97,	1.51) &   &	0.10 (-0.08, 0.28) &  \\
        \hspace{3mm} Chile  &	1.08 (0.85,	1.37) &   &	0.08 (-0.11, 0.27) &  \\
        \hspace{3mm} Mexico  &	0.58 (0.45,	0.75) &   &	-0.54 (-0.73, -0.35) &  \\

        \textbf{Route of infection} &  & 0.563 &  & 0.197\\
        \hspace{3mm} Homosexual/Bisexual (ref) & 1 & & 1 & \\
        \hspace{3mm} Heterosexual & 0.99 (0.81, 1.21) &  & -0.04 (-0.20, 0.11) &   \\
        \hspace{3mm} Other/Unknown & 0.84 (0.61, 1.17) &  & -0.24 (-0.51, 0.02) &  \\
        
        \textbf{Prior AIDS event} & & 0.009 & & ${<}0.001$\\
        \hspace{3mm} No (reference) & 1 & & 1 & \\
        \hspace{3mm} Yes & 1.27	(1.06,	1.52) & & 0.25	(0.1, 0.39) & 	\\
        
        \textbf{Baseline CD4} & \multirow{2}{*}{ 1.13	(1.12,	1.15)} & 	\multirow{2}{*}{${<}0.001$} & \multirow{2}{*}{ 0.10	(0.09, 0.11)} & 	\multirow{2}{*}{	$<0.001$ } \\
        (per 1 square root cells/$\mu$L) &  \\
        
        \textbf{Baseline VL} & \multirow{2}{*}{1.04 (0.95,	1.14) } & 	\multirow{2}{*}{0.357} & \multirow{2}{*}{0.35 (0.27, 0.42)} & 	\multirow{2}{*}{	$<0.001$ } \\
        (per 1 $\log_{10}$ copies/mL) &  \\
       
        \textbf{ART regimen} & & ${<}0.001$  & & ${<}0.001$ \\
        \hspace{3mm} NNRTI-based (reference) &	1 & & 1 & \\
        \hspace{3mm} INSTI-based & 0.44 (0.27, 0.72) &  & -0.60 (-0.93, -0.28) &  \\
        \hspace{3mm} PI-based & 0.95 (0.77, 1.16) &  & 0.07 (-0.10, 0.23) &  \\
        \hspace{3mm} Other & 3.58 (1.49, 8.60) & & 1.06 (0.30, 1.82) &  \\
        
        \textbf{Months to VL measure} & 1.01	(1.00, 1.03) & 0.068 & -0.03	(-0.06, 0.01) & 	0.140 \\
        
        \textbf{Calendar year} & 0.86  (0.84, 0.87) & ${<}0.001$ & -0.12 (-0.13, -0.10) & ${<}0.001$\\
        \hline
    \end{tabular}
\end{table}

 Figure \ref{fig:splines_all} plots the estimated conditional quantiles and probabilities of the outcome being above 1000 and 20. The conditional median and its confidence intervals all fall into the smallest level ``${<}20$''. For age, the general trend is the same as in the model without splines but for patients younger than 40 years old, age increase is correlated with a slight increase in 6-month viral load. The estimated conditional quantiles and probabilities as a function of prior AIDS events shown in the third row of  Figure \ref{fig:splines_all} are very similar to the results in the model without splines. In the second row of  Figure \ref{fig:splines_all}, patients with large baseline viral load values tend to have much higher 6-month viral load measures. 
 
 Table \ref{tab:res_splines} shows the odds ratio and 95\% confidence intervals using splines for all predictors. The p-values in the table are based on the likelihood ratio tests.

Results for analyzing the data using logistic regression and the full likelihood-based model are shown in Table \ref{tab:logistic}.

\section{Simulation Results}

We compared CPMs with some widely used methods for handling DLs under under scenario 2 with $n=1000$ described in Section 4.1, specifically single imputation with $l/2$, single imputation with $l/\sqrt{2}$, multiple imputation, and fully parametric maximum likelihood estimation (MLE). Results are in Table \ref{tab:comparison_mul}.
 
\begin{table}
    \centering
    \caption{Comparison of methods under correct and incorrect model specifications}
    \renewcommand{\arraystretch}{0.53} 
    \begin{tabular}{ccccccc}
    \hline
   Method & Truth & Bias(\%) & Empirical SE & RMSE & Coverage\\
  \hline
   \multicolumn{6}{c}{\textbf{Correct model specification}} \\
   \hline
   \textbf{CPM} & & & & &  \\
   $\beta$ & 1 & 0.258 & 0.040 & 0.040 & 0.951\\
   $Q(0.5|X=0)$ & 1 & 0.226 & 0.045 & 0.045 & 0.953\\
   $Q(0.5|X=1)$ & 2.718 & 0.490 & 0.155 & 0.156 & 0.949\\
   \hline
    \textbf{Single imputation with $dl/\sqrt{2}$} & & & &\\
    $\beta$ & 1 & -10.294 & 0.028 & 0.107 & 0.057\\
    $Q(0.5|X=0)$ & 1 & 3.614 & 0.039 & 0.053 & 0.859\\
    $Q(0.5|X=1)$ & 2.718 & -4.580 & 0.144 & 0.190 & 0.883\\
    \hline
    \textbf{Single imputation with $dl/2$} & & & & \\
    $\beta$ & 1 & -4.247 & 0.030 & 0.052 & 0.732\\ 
    $Q(0.5|X=0)$ & 1 & 0.620 & 0.039 & 0.040 & 0.955\\
    $Q(0.5|X=1)$ & 2.718 & -1.700 & 0.147 & 0.155 & 0.949\\
    \hline
    \textbf{Multiple imputation} & & & & \\
    $\beta$ & 1 & 0.269 & 0.035 & 0.036 & 0.964\\
    $Q(0.5|X=0)$ & 1 & 1.341 & 0.039 & 0.041 & 0.947\\
    $Q(0.5|X=1)$ & 2.718 & -1.576 & 0.148 & 0.154 & 0.945\\
    \hline
    \textbf{MLE} & & & & \\
    $\beta$ & 1 & 0.031 & 0.035 & 0.035 & 0.952\\
    $Q(0.5|X=0)$ & 1 & -0.006 & 0.032 & 0.032 & 0.952\\
    $Q(0.5|X=1)$ & 2.718 & 0.075 & 0.123 & 0.123 & 0.952\\
    \hline 
    \multicolumn{6}{c}{\textbf{Model misspecification}} \\
    \hline
    \textbf{CPM} & & & & &\\

    $\beta$ & 1 & 0.257 & 0.040 & 0.040 & 0.951\\
    $Q(0.5|X=0)$ & 24.997 & 0.070 & 0.444 & 0.444 & 0.953\\
    $Q(0.5|X=1)$ & 35.996 & 0.127 & 0.680 & 0.681 & 0.949\\
    \hline
    \textbf{Single imputation with $dl/\sqrt{2}$} & & & &\\
    $\beta$ & 1 & -61.821 & 0.011 & 0.618 & 0.000\\
    $Q(0.5|X=0)$ & 24.997 & -3.568 & 0.389 & 0.973 & 0.405\\
    $Q(0.5|X=1)$ & 35.996 & -0.818 & 0.665 & 0.727 & 0.935\\
    \hline
    \textbf{Single imputation with $dl/2$} & & & & \\
    $\beta$ & 1 & -55.739 & 0.014 & 0.558 & 0.000\\
    $Q(0.5|X=0)$ & 24.997 & -4.833 & 0.445 & 1.287 & 0.262\\
    $Q(0.5|X=1)$ & 35.996 & 0.240 & 0.698 & 0.703 & 0.887\\
    \hline
    \textbf{Multiple imputation} & & & & \\
    $\beta$ & 1 & -59.690 & 0.016 & 0.597 & 0.000\\
    $Q(0.5|X=0)$ & 25.000 & -3.040 & 0.382 & 0.851 & 1.000\\
    $Q(0.5|X=1)$ & 35.996 & -1.014 & 0.660 & 0.754 & 1.000\\
    \hline
    \textbf{MLE} & & & & \\
    $\beta$ & 1 & -62.108 & 0.012 & 0.621 & 0.000\\
    $Q(0.5|X=0)$ & 24.997 & -4.838 & 0.303 & 1.247 & 0.021\\
    $Q(0.5|X=1)$ & 35.996 & -3.468	& 0.535 & 1.358 & 0.451\\
    \hline
    \end{tabular}
    \label{tab:comparison_mul}
\end{table}

We show the performance of CPMs with moderate and severe link function misspecification in Table \ref{tab:mis}. 

\begin{table}
    \centering
    \caption{Simulation results for link function misspecification with one lower DL at 0.25}
    \label{tab:mis}
    \renewcommand{\arraystretch}{0.53} 
    \begin{tabular}{@{\extracolsep{5pt}}cccccccc@{}}
    \hline
    & & \multicolumn{3}{c}{n=100} & \multicolumn{3}{c}{n=500}\\
    \cline{3-5} \cline{6-8}
  Parameter & Truth & Bias(\%) & RMSE & Coverage  &   Bias(\%)
  & RMSE & Coverage\\
  \hline
   \textbf{Logit Link} & & & & & & & \\
    $Q(0.5|X=0)$ & 1 &  0.006 & 0.142 & 0.949 & 0.287 & 0.064 & 0.942\\
    $Q(0.5|X=1)$ & 2.718 & 2.827 & 0.514 & 0.947 & 1.965 & 0.233 & 0.936\\
    $F(1.5|X=0)$ & 0.658 & 0.923 & 0.057 & 0.944 & 0.869 & 0.024 & 0.940\\
    $F(1.5|X=1)$ & 0.276 & -5.221 & 0.062 & 0.939 & -4.862 & 0.030 & 0.912\\
  \hline
  \textbf{Loglog Link} & & & & & & & \\
    $Q(0.5|X=0)$ & 1 & -4.116 & 0.232 & 0.953 & -3.241 & 0.042 & 0.932 \\
    $Q(0.5|X=1)$ & 2.718 & -10.673 & 0.491 & 0.887 & -11.861 & 0.192 & 0.603\\
    $F(1.5|X=0)$ & 0.658 & -0.4602 & 0.045 & 0.961 & -1.081 & 0.020 & 0.947 \\
    $F(1.5|X=1)$ & 0.276 & 8.092 & 0.076 & 0.900 & 10.164 & 0.010 & 0.809\\
    
    \hline 
    \end{tabular}
\end{table}

The simulation results for multiple detection limits are in Table \ref{tab:mul_dl}. 

\begin{table}
\begin{threeparttable}
    \centering
    \caption{Simulation results for multiple DLs}
    \label{tab:mul_dl}
    \renewcommand{\arraystretch}{0.53} 
    \begin{tabular}{@{\extracolsep{5pt}}ccccccccc@{}}
    \hline
    & \multicolumn{4}{c}{n=50 for each site} & \multicolumn{4}{c}{n=300 for each site}\\
     \cline{2-5}\cline{6-9}
  Parameter & Bias(\%) & Bias & RMSE & Coverage  &   Bias(\%) & Bias
  & RMSE & Coverage\\
  \hline
   \textbf{Scenario 1} & & & & & & \\
    $\beta$ & 1.871 & 0.019 & 0.111 & 0.948 &  0.249 & 0.003 & 0.044 & 0.948 \\
    $Q(0.5|X=0)$ & -0.071 & -0.001 & 0.12 & 0.948 & 0.001 & -0.001 & 0.047 & 0.953\\
    $Q(0.5|X=1)$ & 1.137 & 0.031 & 0.404 & 0.952 & 0.042 & 0.001 & 0.164 & 0.947 \\
    $F(1.5|X=0)$ & 0.258 & 0.002 & 0.044 & 0.950 & 0.134 & 0.001 & 0.017 & 0.954 \\
    $F(1.5|X=1)$ & -1.115 & -0.003 & 0.050 & 0.950 & -0.435 & -0.001 & 0.020 & 0.952\\
    \hline
    \textbf{Scenario 2} & & & & & & \\
    $\beta$ & 3.706 & 0.037 & 0.186 & 0.948 & 0.455 & 0.005 & 0.068 & 0.954\\
    $Q(0.03|X=0)$ & 3.440 & 0.005 & 0.037 & 0.953 & 0.152 & 0.000 & 0.014 & 0.952\\
    $Q(0.03|X=1)$ & -1.952 & -0.008 & 0.071 & 0.953 & 1.795 & 0.007 & 0.046 & 0.9545\\
    $F(0.05|X=0)$ & 11.276 & 0.000 & 0.000 & 0.974 & -2.917 & 0.000 & 0.000 & 0.953\\
    $F(0.05|X=1)$ & 79.059 & 0.000 & 0.000 & 0.961 & -3.762 & 0.000 & 0.000 & 0.944\\
    \hline
    \textbf{Scenario 3} & & & & & & \\
    $\beta$ & 1.891 & 0.019 & 0.106 & 0.958 & 0.248 & 0.003 & 0.041 & 0.962\\
    $Q(0.5|X=0)$ & -0.202 & -0.002 & 0.117 & 0.964 & -0.092 & -0.001 & 0.047 & 0.969 \\
    $Q(0.5|X=1)$ & 0.939 & 0.026 & 0.395 & 0.950 & 0.041 & 0.001 & 0.160 & 0.949 \\
    $F(1.5|X=0)$ & 0.223 & 0.002 & 0.045 & 0.963 & 0.159 & 0.001 & 0.017 & 0.962 \\
    $F(1.5|X=1)$ & -1.221 & -0.003 & 0.050 & 0.948 & -0.383 & -0.001 & 0.020 & 0.955\\
    \hline
    \textbf{Scenario 4} & & & & & & \\
    $\beta$ & 1.926 & 0.019 & 0.175 & 0.945 & 0.206 & 0.002 & 0.065 & 0.954\\
    $Q(0.03|X=0)$ & 4.032 & 0.006 & 0.039 & 0.955 & 0.332 & 0.001 & 0.014 & 0.951\\
    $Q(0.03|X=1)$ & 7.982 & 0.033 & 0.125 & 0.950 & 2.096 & 0.009 & 0.049 & 0.950\\
    $F(0.05|X=0)$ & 21.028 & 0.000 & 0.000 & 0.965 & -1.349 & 0.000 & 0.000  & 0.952\\
    $F(0.05|X=1)$ & 110.371 & 0.000 &  0.000 & 0.956 & -0.940 & 0.000 & 0.028 & 0.943\\
    \hline
    \textbf{Scenario 5} & & & & & & \\
    $\beta$ & 1.838 & 0.018 & 0.111 & 0.957 & 0.250 & 0.003 & 0.044 & 0.960\\
    $Q(0.5|X=0)$ & -0.046 & -0.001 & 0.115 & 0.948 & 0.440 & 0.004 & 0.047 & 0.951\\
    $Q(0.5|X=1)$ & 2.019 & 0.054 & 0.412 & 0.968 & 0.171 & 0.005 & 0.165 & 0.963\\
    $F(1.5|X=0)$  & 0.406 & 0.003 & 0.04 & 0.945 & -0.059 & -0.000 & 0.229  & 0.960\\
    $F(1.5|X=1)$ & -0.059 & -0.000 & 0.050 & 0.960 & -0.319 & -0.001 & 0.020 & 0.961\\
    \hline
    \end{tabular}

\begin{tablenotes}\footnotesize
\item The percent bias in scenario 4 is relatively high due to the small true values.
\end{tablenotes}
\end{threeparttable}
\end{table}